\documentclass[]{article}
\usepackage{caption}
\usepackage{amsfonts}
\usepackage[slantedGreek]{mathpazo}
\usepackage{graphicx,float,array}
\usepackage{amsmath,amssymb,amsthm,hyperref}
\usepackage[margin=1in]{geometry}
\usepackage{multirow}
\usepackage{hyperref}

\usepackage{graphicx,float}
\usepackage{algorithm}
\usepackage{algpseudocode}
\usepackage[utf8]{inputenc} 
\usepackage{caption}
\usepackage{marginnote}
\usepackage{xcolor}
\usepackage[numbers]{natbib}

\newcommand{\defeq}{\mathrel{\mathop:}=}
\usepackage[normalem]{ulem}              
\usepackage{setspace}                    
\usepackage[colorinlistoftodos]{todonotes}
\usepackage{algorithm}
\usepackage{algpseudocode}
\usepackage[scr=boondoxo]{mathalfa}
\usepackage{caption}
\usepackage[slantedGreek]{mathpazo}
\usepackage{graphicx,float,array}
\usepackage{amsmath,amssymb,amsthm,hyperref}
\usepackage[margin=1in]{geometry}
\usepackage{multirow}
\usepackage{hyperref}
\usepackage[utf8]{inputenc} 
\usepackage{url}
\usepackage{textcomp}

\DeclareMathOperator*{\argmax}{arg\,max}
\DeclareMathOperator*{\argmin}{arg\,min}

\makeatletter
\renewcommand\paragraph{\@startsection{paragraph}{4}{\z@}%
	{-2.5ex\@plus -1ex \@minus -.25ex}%
	{1.25ex \@plus .25ex}%
	{\normalfont\normalsize\bfseries}}
\makeatother
\graphicspath{{./fig/}}

\begin{document}	
\title{Bayesian Calibration for Activity Based Models}
\author{
	Laura Schultz, Joshua Auld and Vadim Sokolov\thanks{Vadim Sokolov is Assistant Professor in Operations Research at George Mason University. email:vsokolov@gmu.edu}}
\date{First Draft: December, 2019\\This Draft September 3, 2022}
\maketitle
%
\begin{abstract}
We consider the problem of calibration and uncertainty analysis for activity-based transportation simulators. Activity-Based Models (ABMs) rely on statistical modeling of individual travelers' behavior to predict higher-order travel patterns in metropolitan areas. Input parameters are typically estimated  from traveler surveys using maximum likelihood. We develop an approach that uses a Gaussian Process emulator to calibrate those parameters using traffic flow data. Our approach extends traditional emulators to handle the high-dimensional and non-stationary nature of transportation simulators. We introduce a deep learning dimensionality reduction model that is jointly estimated with Gaussin Process model to approximate the simulator.  We demonstrate the methodology using several simulated examples as well as by calibrating key parameters of the Bloomington, Illinois model.
\end{abstract}

\section{Introduction}\label{sec:intro}

Transportation activity-based simulators (ABMs) represent an individual traveler's activity patterns and trips throughout the day by using nested choice models. The generated trips are then simulated in a traffic flow simulator to learn system-level patterns. These behaviorally-realistic models require a high-resolution representation of network flows and, thus, are computationally expensive. The very same flexibility which makes these simulation models appealing, also makes their calibration problems intractable, with the number of simulations required to find an optimal solution growing exponentially as the input dimension increases~\cite{shan2010survey,more_benchmarking_2009}. As a result, the use of these simulators is currently limited to what-if analysis. 

This paper focuses on calibrating the static choice model parameters used in activity-based simulators. The goal of calibration is to find values of the simulator's input parameters $\theta$ that minimizes the deviance between observed data and simulator's outputs. We assume that observed $y(z)$ and the simulated $\eta(z,\theta)$ traffic flows are related via the following equation:

\begin{equation}
y(z) =\eta \left(z,\theta \right)+\epsilon(\theta) + e,
\end{equation}\label{eq:y=eta+ep+e}
where $z$ is a set of observed inputs the modeler is certain about (e.g. day of the week) and $\theta$ is a set uncertain inputs (e.g. value of traveler's time) that need to be calibrated; $\epsilon(\theta)$ captures any structural inadequacy of the simulator; and ${e}$ represents the observation error and residual variation inherent to the true process which generates $y$.


The problem of simulator calibration is well studied in the transportation literature~\cite{balakrishna_offline_2003}.  However, research has primarily focused on traditional, aggregated models, which use origin-destination estimation to represent demand~\cite{cascetta_unified_1988,zhou_structural_2007,zhou_dynamic_2004}. The typical approach has been to use optimization methods~\cite{zhang_lei_integrating_2013,zhang_efficient_2017}; specifically, by using simulation-based optimization approaches which treat the simulator as a black-box  \cite{stevanovic_stochastic_2008,mesbah_optimization_2011-1,baldi_simulation-based_2019,brian_park_stochastic_2009}, such as Genetic Algorithm (GA)~\cite{cheu_calibration_1998,ma_genetic_2002,mesbah_optimization_2011}, Simultaneous Perturbation Stochastic Approximation (SPSA)~\cite{lu_enhanced_2015,cipriani_gradient_2011,lee_new_2009}, and exhaustive evaluation~\cite{hale_optimization-based_2015}, with relative success~\cite{yu_calibration_2017}. However, due to the non-linear, non-stationary, and stochastic nature of new ABM simulators, straightforward application of existing optimization techniques have limited applicability given they require large number of simulation runs \cite{florian_hybrid_2001}.

Some alternatives substitute a complex simulator with a set of lower-fidelity simulators to approximate the ABM's solution \cite{chong_simulation-based_2017,osorio_simulation-based_2017} or through hierarchical decomposition to more manageable sub-problems \cite{browning_applying_2001,michelena_optimal_2019,baldi_simulation-based_2019}. However, while these multi-fidelity approaches do improve on the problem of costly run times, the number of samples required still remains high and the curse of dimensionality prevents us from using these methods in high-dimensional problems. For example, if a $20$-dimensional simulator's run could be reduced to one second, it would still take 3.171 trillion years of computing ($10^{20}$ runs) to evaluate solutions on a grid with just $10$ points per dimension. Further, additional development of these lower-fidelity models would be necessary and results in significantly increased modeling efforts and resources, e.g. modeler's effort and budgets. Divide-and-conquer methods, or parallelization-based techniques, have also been studied to address the dimensional issue~\cite{garcia_decentralized_2007,zhang_lei_integrating_2013,wu_parallel_2016,hutchison_parallel_2013,desautels_parallelizing_2014,shah_parallel_2015,wang_parallel_2016} but can only provide a multiplicative expansion to an exponential growth in sampling requirements. 

Another set of techniques relies on Bayesian updating to solve the calibration problem \cite{hazelton_statistical_2008,flotterod_general_2010,flotterod_bayesian_2011,zhu_calibrating_2018}. A probabilistic surrogate function is automatically constructed to emulate the simulation's behavior without the additional modelling development of previous methods. The integration of available data allows the non-parametric estimate to recommend optimal samples sequentially~\cite{gramacy_adaptive_2009,snoek_input_2014,danielski_gaussian_2013,rasmussen_gaussian_2006,romero_navigating_2013}, which makes the approach sample-efficient~\cite{brochu_tutorial_2010,frazier_bayesian_2018,osorio_simulation-based_2017,gramacy_adaptive_2009,higdon_computer_2008,taddy_big_2018,snoek_practical_2012,snoek_scalable_2015,rosipal_kernel_2003}. We propose to use this Bayesian surrogate-based approach to automatically construct a low-fidelity surrogate to solve the calibration problem. We use a non-parametric Gaussian Process (GP) as our surrogate \cite{sacks_design_1989} and expand the calibration framework proposed by \cite{kennedy_bayesian_2001, schultz_bayesian_2018, sha2020applying}.

However, in order to apply the algorithm to our transportation simulators, two drawbacks of the GP surrogate must be addressed. First, it assumes that the underlying function to be approximated is smooth \cite{shan_survey_2010,donoho_high-dimensional_2000}, which our ABMs are not. Some approaches which assume a specific functional form of the simulator were proposed to tackle the non-smooth nature of the problem  \cite{gramacy_adaptive_2009,schmidt_considering_2011,binois_practical_2018,schmidt_spatiotemporal_2017} but require prior knowledge to choose the right form. Second, while capable of reproducing a wide range of behaviors and providing a highly flexible fit~\cite{gramacy_particle_2011}, the non-parametric nature of GPs also hinders success in high-dimensional settings without large sample sizes. As a result, GP modeling, uncertainty analysis, and optimization applications have historically been restricted to characteristically low-dimensional problems~\cite{shorter_efficient_1999,constantine_active_2015,bates_experimental_1996,booker_rigorous_1998,koch_statistical_1999,srivastava_method_2004,tu_variable_2003}; the curse of dimensionality problem continues to be unsolved~\cite{shan_survey_2010}. We address both of those drawbacks by introducing a deep learning pre-processing step that reduces dimensionality and changes the nature of the simulator function. 

Previously proposed dimensionality reduction techniques make specific assumptions, e.g. that simulator function is additive and exploit this structure~\cite{duvenaud_automatic_2014, durrande_additive_2012,kandasamy_bayesian_2015,ma_dimensionality-driven_2018,muehlenstaedt_data-driven_2012,binois_quantifying_2015,constantine_active_2015,babichev_slice_2018}. Dimensionality Reduction seeks influential substructures to obtain a low dimensional input that is a projection of the original high-dimensional one. Dimensionality Reduction techniques~\cite{nagel_agent-based_2012,adragni_sufficient_2009,glaws_inverse_2017} include variable screening~\cite{marrel_global_2011,chen_high-fidelity_2015,saheb_ettabaa_adaptive_2018,naik_constrained_2005}, kernel construction~\cite{liu_dimension_2017, cortes_rational_2004,wilson_deep_2016,durrande_additive_2012,hinton_using_2008}, Partial Least Squares (PLS), Principal Component Analysis (PCA), and sliced inverse regression (SIR) \cite{djukic_efficient_2012}.

In essence, our contribution lies in developing a surrogate model that combines the non-linear deterministic dimensionality reduction technique with the probabilistic GP model and developing an inference scheme that jointly estimates parameters of those two components. We then apply the sequential design experiment techniques along with the proposed surrogate model to implement Bayesian optimization for an activity-based transportation simulator. We show that our proposed surrogate outperforms ordinary GPs and more complex GPs which are designed to deal with non-stationary data. 







The rest of the paper is organized as follows: in Section \ref{sec:polaris}, we detail the typical structure of transportation simulators and the paper's example ABM POLARIS~\cite{auld_polaris:_2016}; Section \ref{sec:abmruns} provides exploratory analysis of the simulator to motivate our methodology. Section \ref{sec:model} outlines the methodological contribution of this paper; Section \ref{sec:results} demonstrates our methodology on the problem of calibrating our example model; and Section \ref{sec:discussion} discusses avenues for future research.

\section{POLARIS: an Agent-Based Model}\label{sec:polaris}

\begin{figure}[H]
	\centering
	\includegraphics[width=0.6\linewidth]{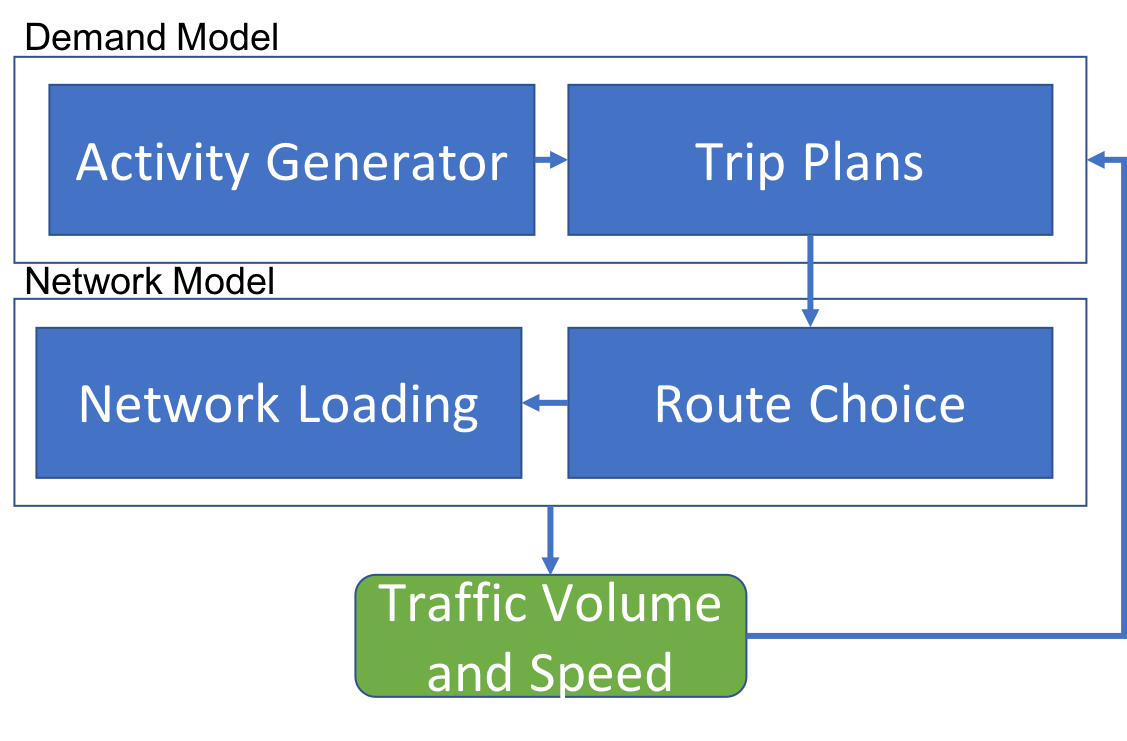}
	\caption{An example of simulation flow diagram for an integrated transportation simulator}
	\label{fig:polaris}
\end{figure}

An activity-based transportation simulator has two components: network and demand. The demand simulator includes several inputs, such as land-use variables, socio-demographic characteristics of the population, road infrastructure, and congestion patterns. Then the activity-based models generate the trip tables. Parameters of statistical models that generate the trips are typically estimated using small-sample surveys. Inadequacy of the models combined with low sample sizes lead to structural errors and large unexplained variance in those models. The network component uses trips generated by the demand model as an input and calculates the resulting congestion patterns. Figure~\ref{fig:polaris} shows an example of the the overall architecture.

Typically, the demand-network models are executed several times in a loop to achieve an approximation to equilibrium in the network flows. The traffic volume and speed are the main outputs of the simulator that are used in our calibration procedure by comparing with the measured traffic flows. 

\subsection{Demand Module}

A demand module first generates a set of trips for each of the travelers (agents) by first assigning a destination and series of choices to individual users. For POLARIS, a Weibull hazard model is used~\cite{auld_dynamic_2011} to estimate the probability $h_{ij}(t)$ of an activity of type $j$ occurring for traveler $i$ at time $t$:
\begin{equation}
h_{ij}(t) = \gamma_{j} t^{\gamma-1}\exp(-\beta_j^Tx_i),
\end{equation}

where $\gamma_j$ is the shape parameter of the Weibull distribution, $x_i$ are observed attributes of the traveler and $\beta$ are the model parameters estimated from a survey. 

Next, the order and timing for each activity, such as location, start time, duration and mode, is allocated by the trip planning model which is a set of discrete choice models; POLARIS uses the multinomial logit model~\cite{ben-akiva_discrete_1985} to choose the destinations for each given activity type. This model is derived from random utility maximization theory, which states that for each decision maker $n$ and location $i$, there is a utility $U_{in}$ associated with selecting location $i$. A generalized linear model for utility is assumed

\begin{equation}
U_{in} = \beta^Tx_{ni} + \epsilon_{in},
\end{equation}

where $\beta_i$ represent the parameters estimated from survey data; $x_{ni}$ are observed attributes of traveler $n$ and location $i$, such as travel time to, and land use attributes of, $i$; and $\epsilon_{in}$ is an unobservable random error which follows a Gimbel distribution so that the probability of traveler $n$ choosing destination $i$ is given by

\begin{equation}
	p_{ni} = \dfrac{\exp(V_{in})}{\sum_{j \in C_n} \exp(V_{jn})}.
\end{equation}
Here $V_{in} = \beta^Tx_{ni}$.  The choice set $C_n$ is formed using a space-time prism constraint \cite{auld_polaris:_2016}.

Once attributed, the planning order is determined using the Ordered Probit model, which is an extension of the basic probit model for binary responses extended to ordinal responses $y \in \{1,\ldots K\}$. Under this model, the probability of $y$ being equal to $k$ is given by
\[
p(y = k) = p(l \le \alpha_k) - p(l\le \alpha_{k-1}),
\]
where $\alpha_1,\ldots\alpha_{K-1}$ is a set of threshold-values, and $l = \beta^Tx + \epsilon,~~~\epsilon\sim N(0,\Sigma)$.

\subsection{Network Module}

Once the trips, defined by the origin, destination, departure time and mode, are generated, the network model performs Dijkstra's algorithm to determine route on the transportation network. Once these choices on routes are selected, destination and departure times are assigned to each agent and the resulting congestion levels within the road network is produced.  Kinematic Wave theory of traffic flow~\cite{newell_simplified_1993} is used to simulate traffic flow on each road segment. This model has been recognized as an efficient and effective method for large-scale networks~\cite{lu_dynamic_2013} and dynamic traffic assignment formulations. 

In particular, the Lighthill-Whitham-Richards (LWR)  model ~\cite{lighthill_kinematic_1955,richards_shock_1956} along with discretization scheme proposed by Newell~\cite{newell_simplified_1993} is used. The LWR model is a macroscopic traffic flow model. It is a combination of a conservation law defined via a partial differential equation and a fundamental diagram. The fundamental diagram is a flow-density relation $q(x,t) = q(\rho(x,t))$.  The nonlinear first-order partial differential equation describes the aggregate behavior of drivers. The density $\rho(x,t)$ and flow $q(x,t)$, which are continuous scalar functions,  satisfy the equation 
\begin{equation*}\label{eqn:lwr1}
\frac{\partial\rho(x,t)}{\partial t} + \frac{\partial q(x,t)}{\partial x} = 0.
\end{equation*}
This equation is solved numerically by discretizing  time and space. 

In addition, intersection operations are simulated for signal controls, as well as stop and yield signs. The input for LWR model is a fundamental diagram that relates flow and density of traffic flow on a road segment. The key parameter of fundamental diagram is critical flow, which is also called road capacity. The capacity is measured in number of vehicles per hour per lane, and provides theoretical maximum of flow that can be accommodated by a road segment.

\section{Bayesian Model Formulation}\label{sec:model}

Our calibration problem is formulated as an optimisation problem with the objective of minimizing the difference between the observed traffic flows $y$ and simulated flows $\eta\left(z,\theta\right)$:

\begin{equation}\label{eq:opt}
\theta^\star \in \argmin_{\theta \in A}\ \ \mathcal{L}\left(y, \eta \left(z, \theta \right)\right) \\
\end{equation}

where the vector of input parameters $\theta \in \mathbb{R}^p$ represent coefficients of choice models used by the transportation ABMs and the constraint set $A\in \mathbb{R}^{p\times 2}$ represents a modeler's prior knowledge about feasible ranges of the parameters $a_{i1}\le \theta_i \le a_{i2}$. The univariate loss function $\mathcal{L}$ calculates how close the observed data $y$ is to the simulated outputs $\eta \left(z, \theta \right)$. Without loss of generality, we omit the static variable $z$ that is known to the modeler and does not need to be calibrated.

The form of the loss function $\mathcal{L}$ depends on the observation error $e$, and systematic error $\epsilon(\theta)$ terms of Equation \ref{eq:y=eta+ep+e}. Assuming both of the error terms to be Normal with zero mean, we get the least-squares loss function 
\[
\mathcal{L}(y,\eta (\theta)) = \lVert y - \eta \left(\theta \right) \rVert_2^2
\]

For brevity, we will refer to the loss function as $\mathcal{L}(\theta)$ going forward. 

Because our application involves an ABM simulator $\eta$ which is computationally expensive and has only a limited number of simulations which can be carried out, we use the Bayesian Optimisation (BO) approach for model calibration. 

We start by assuming that the loss function is continuous and placing a prior over continuous functions defined on $[0,\infty]$. Then, we compute an ensemble of initial simulator runs at $N$ input setting $\{\theta^{(i)},\mathcal{L}(\theta^{(i)})\}_{i=1}^N$. These values $[\theta^{(1)},\ldots,\theta^{(N)}]$ are generated using Latin hypercube sampling \cite{mckay_comparison_1979}. With the set of input-output pairs, a Bayesian statistical model is formulated and integrate with the data to calculate its posterior. Each additional sample recommended for evaluation is then iteratively chosen using the sequential design of experiment paradigm; the loss function is evaluated according to a chosen heuristic to determine the optimal next sample which has the greatest potential for improving the estimated solution~\cite{mockus_bayesian_2012}.

We use a Gaussian Process (GP) to model the value of the loss function $\mathcal{L}(\theta)$ at new locations and to perform the sequential design of experiment. One key component of our formulation is the use of dimensionality reduction pre-processing to accelerate the search for optimal location. More specifically, we use deep learning, non-linear dimension reducer. In this section we describe the dimensionality reduction and Bayesian optimisation techniques we propose to solve the problem of calibrating a high-dimensional transportation simulator.





\subsection{Exploratory Analysis of POLARIS Simulator Runs}\label{sec:abmruns}

We use $d=5$ input parameters $\theta = \left[\theta_1,\cdots,\theta_5\right]$, described in Table \ref{tab:params}, in order to demonstrate our framework. These parameters were identified by one of the model developers as the most important set to be calibrated. 

\begin{table}[H]
    \centering
    \begin{tabular}{p{30mm}|l|p{65mm}|l}
	Component &  Name & Description & Range\\\hline
    Mode Choice & \verb|HBO_B_male_taxi| & Effect of traveler's sex on taxi utility & [$0.298$, $2.298$]\\
    Mode   Choice& \verb|NHB_B_dens_bike| & Effect of population density on bike utility & [$6.601$, $8.601$]\\
    Mode   Choice& \verb|HBO_ASC_TAXI| &  Intercept in taxi utility
 & [$2.34$, $4.34$]\\
    Destination   Choice& \verb|THETAR_WORK| & Effect of retail accessibility
    & [$-8.553$, $-6.553$]\\
    Destination Choice& \verb|GAMMA_SERVICE| & Distance decay & [$7.038$, $9.038$]
    \end{tabular}
    \caption{Activity-Based model parameters and their lower and upper bounds}
    \label{tab:params}
\end{table}

For each input parameter setting, sum of  simulated turn delayes, averaged over $5$-minute intervals for the $24$ hours period are taken as the simulator outputs. The resulting $288$ outputs are used to calculate the mean squared error objective function $\mathcal{L}(\theta)$ between simulated and observed values:
\begin{equation}
\mathcal{L}(\theta)= \frac{1}{288}\sum_{i=1}^{288}(y_i - \eta(\theta)_i)^2.
\end{equation}

To explore the nature of the input-output relationships, we use the Morris screening technique~\cite{saltelli_sensitivity_2004}, which is widely used for high-dimensional models, to explore potential non-linearity and any interactions among the input parameters $\theta$.  The Morris procedure starts by generating an initial set of origin samples $\theta^{(1)},\ldots,\theta^{(r)}$. Each variable in $\theta^{(i)}$ is uniformly sampled from its corresponding range. A set of $r$ trajectories are then built, where each trajectory starts at point $\theta^{(i)}$ and is generated by perturbing each of the $d$ variables one at a time by a fixed value $\delta$. For our analysis, we use $r=6$ trajectories constructed by $d=5$ steps, thus resulting in a sample of $N=36$ input-output pairs. 

Given this initial sample set, we calculate the finite difference, called the Elementary Effect (EE), for each trajectory $i$ and each variable $j$:

\begin{equation}
    EE_{i,j} = \frac{\mathcal{L}(\theta_1^i,\theta_2^i,\cdots, \theta_{j}^i+ \delta,\cdots,\theta_d^i) - \mathcal{L}(\theta_1^i,\theta_2^i,\cdots,\theta_{j}^i,\cdots,\theta_m^i)}{\delta}
\end{equation}

Then Morris procedure analyses elementary effects using three statistics
\begin{equation}
\mu_j = \frac{1}{r} \sum^r_{i=1} EE_{i,j},\quad \sigma_j = \sqrt{\frac{1}{r} \sum_{i=1}^r(EE_{i,j} - \mu_j)^2},\quad \mu_j^\star = \frac{1}{r} \sum^r_{i=1} \vert EE_{i,j} \vert.
\end{equation}

Where $\mu_j$ measures the average, overall influence of the input $j$ on the output; $\sigma_j$ measures the aggregate degree of higher-order interactions between the $i$th factor and all other $i \ne j$ input factors, known as the standard deviation of the EE. Finally, $\mu^\star_i$  is an adjusted version of $\mu_j$ to handle cases when positive and negative effects may cancel one another in $\mu$ by using the absolute values of EEs~\cite{campolongo_effective_2007}.

Importance of a variable can be found by comparing $\mu$ and $\mu^\star$~\cite{saltelli_sensitivity_2004}. Variables with a high value of $\mu$ suggest a high level of importance and that the effect is monotonic while high values of $\mu^\star$ with low $\mu$ values indicate non-monotonic and/or non-linear behavior. Variables with an equivalent value of $0$ for both $\mu$ and $\mu^\star$ have no effect on the output and can be discarded. A further comparison of $\mu^\star$ with $\sigma$ provides information on non-linear and/or interaction effects. High proportions of $\sigma$ to $\mu^\star$ indicates that the variable interacts with other variables and/or the objective function is  non-linear with respect to this variable. Low proportions indicates no interactions and linearity. 

\begin{figure}[H]
\centering
		\begin{tabular}{cc}
		\includegraphics[width=.4\linewidth]{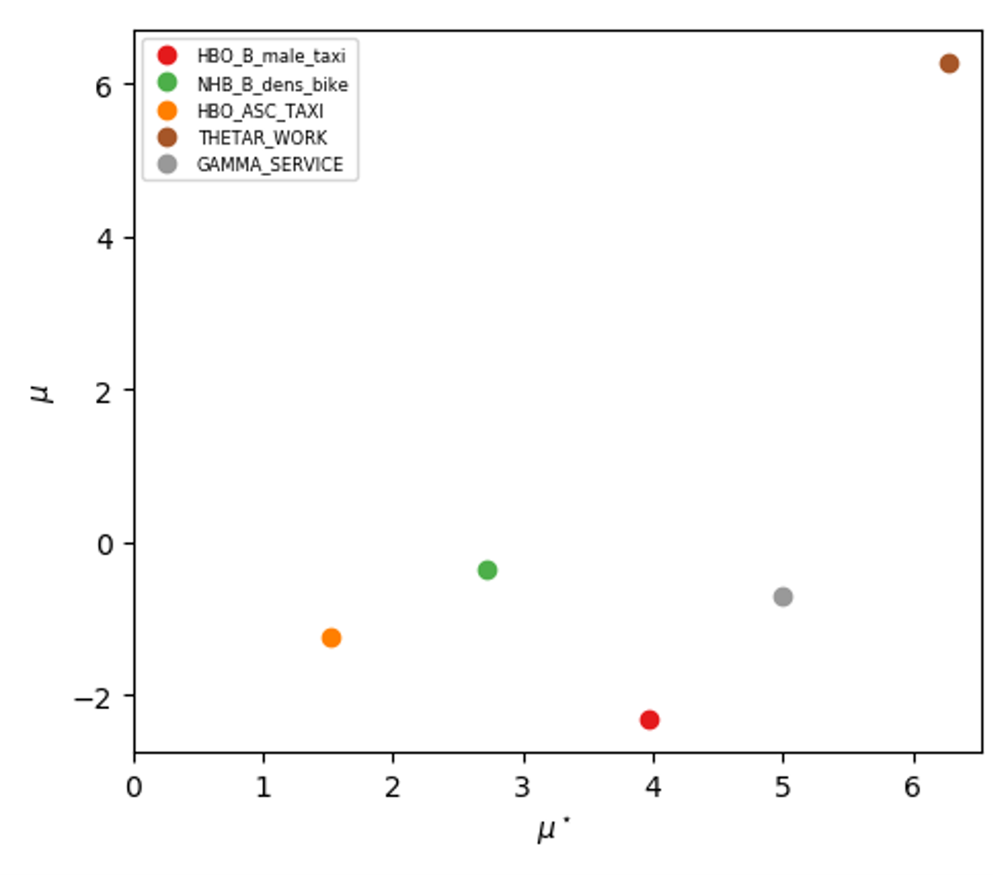} 	&
		\includegraphics[width=.39\linewidth]{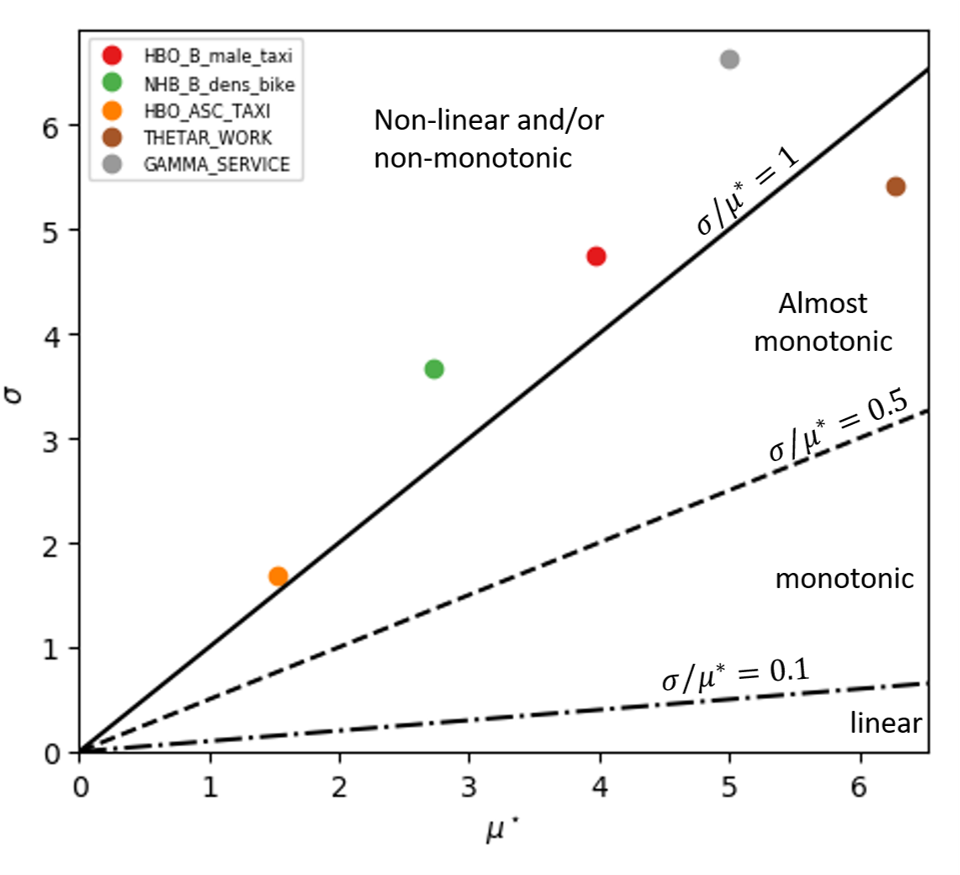}\\
		(a) $\mu$ vs $\mu^\star$ & (b) $\mu^\star$ vs $\sigma$
\end{tabular}
	\caption{(a) A comparison of $\mu$ and $\mu^\star$~\cite{saltelli_sensitivity_2004} for importance ranking. Since no variable has a $\mu^\star=0$, all variables are significant. (b) A comparison of $\mu^\star$ and $\sigma$. High values of $\mu^\star$ with low $\mu$ values indicate non-monotonic and/or non-linear behavior while high values of $\sigma$ indicate interactions.}
	\label{fig:morris}
\end{figure}

Figure \ref{fig:morris}(a) plots the values of calculated statistics for each of our $5$ unknown variables. All five variables have a positive value of $\mu^\star$ and, thus, all five variables have a significant effect on the output. Variables  \verb|GAMMA_SERVICE| and \verb|THETAR_WORK| have the highest importance and influence on the objective function. The value of $\mu$ and $\mu^\star$ are similar for \verb|THETAR_WORK|, suggesting positive monotonic behavior; \verb|HBO_ASC_TAXI| suggests a negative monotonic behavior but the right graph implies an influential non-linear component as well. 

Figure \ref{fig:morris}(b) shows most of the variables have high values of $\sigma$, indicating strong interaction effects and/or non-linear behavior. While \verb|HBO_ASC_TAXI| is comparatively lower, its ratio to $\mu^\star$ remains high at $86\%$ of $\mu^\star$. This indicates a significant interaction or non-linear effect on the other variables but with a lower direct influence on the overall output.

In conclusion, all variables prove important, and the presence of interactions prevent us from using variable screening or shrinkage as a dimensionality reduction technique. Further, the non-linear and non-monotonic nature of the input-output relations advises against using traditional projection-based dimensionality reduction methods, which assume the opposite.

\subsection{Bayesian Optimization}\label{sec:BO}

The Bayesian optimization procedure starts with an initial sample set $\mathcal{D} = \{\theta^{(i)},\mathcal{L}(\theta^{(i)})\}_{i=1}^N$. Two common sample-generating procedures used in computer simulations are the simple random and Latin Hypercube (LHS) sampling techniques~\cite{helton_latin_2003}. Simple random sampling draws a uniformly-distributed random value from each individual variable's domain to construct a sample set; but, though easier to implement, the procedure cannot guarantee that every portion of the subspace is sampled~\cite{helton_latin_2003}. LHS first subsections each variable range into equal intervals and selects a random value within each slice before randomly pairing each variable's new subset to construct a sample~\cite{mckay_comparison_1979}. Our Bayesian optimization procedure uses a LHS configuration to generate our initial sample set.

A surrogate model $F$ is then estimated using a univariate Gaussian Process to quantify the uncertainty at other unsampled points in the domain; a GP is fully characterized by its mean function $m$ and its covariance function, or kernel, $k$~\cite{gramacy_particle_2011}.
\label{key}
\begin{equation}
\mathcal{L}(\theta) \sim \mathcal{GP}(m(\theta),k(\theta,\theta^{\prime}))
\end{equation}
where
\begin{equation}
\begin{split}
m(\theta) &= \mathbb{E}[\mathcal{D} \vert \theta]\\
k(\theta,\theta^\prime) &= \mathbb{E}[(\theta-m(\theta))(\theta^\prime - m(\theta^\prime)]
\end{split}
\end{equation}

The mean function $m(\theta)$ is unrestricted but, in practice, routinely assumed to be zero. The correlation equation $k(\theta,\theta^\prime)$ must result in a positive semi-definite matrix; common choices include the Squared Exponential, which is both stationary and isotropic, and the Mat\'ern covariance function, which provides less smoothness through the reduction of the covariance differentiability. General references can be found in \cite{abrahamsen_review_1997} and \cite{duvenaud_automatic_2014}. 
In this paper, we use the Mat\'ern covariance function:
\[ k_{Matern}(\theta,\theta^{\prime},\Omega) = \frac{{2}^{1-v}}{\Gamma(v)}\left[\frac{\sqrt{2v}|\theta - \theta^{\prime}|}{\lambda}\right]^v K_v\left[\frac{\sqrt{2v}|\theta-\theta^\prime|}{\lambda}\right] \]

where $\Omega = (v,\lambda)$ with lengthscale $v$ and smoothness $\lambda$ as non-negative hyperparameters; $\Gamma$ is the gamma function; and $K_{v}$ is a modified Bessel function.

Additionally, we incorporate the inadequacy errors of $\epsilon(\theta)$ in Equation \ref{eq:y=eta+ep+e} via a specialized Linear ``nugget'' $ r(\theta) = \mathcal{N}(0,I\sigma_{\epsilon}^2)$ ~\cite{gramacy2012cases}:

\begin{equation}
\mathcal{L}(\theta) \sim \mathcal{GP}\left(m(\theta), K(\theta) = k(\theta,\theta^{\prime})+r(\theta)I\right)
\end{equation}
where $I$ represents the identity matrix.

Once conditioned on the observed data $\mathcal{D}$, the final density known as the posterior distribution, is Normal:

\begin{equation}\label{eq:posterior}
\left[y(\theta)  \mid  \mathcal{D}, \theta,\Omega\right] \sim \mathcal{N} \left(\mu, \Sigma\right),
\end{equation} 
where $\Omega$ are the parameters of the kernel function, and with the following summary statistics:
\begin{equation}\label{N_pred}
\begin{split}
\mu &= m(\theta) + k(\theta)(K_{\mathcal{D}})^{-1}(y-m_\mathcal{D})\\
\Sigma &= k(\theta,\theta) + r(\theta)-k(\theta)(K_{\mathcal{D}})^{-1}k(\theta)^T
\end{split}
\end{equation}
where $k(\theta)=\left(k(\theta,\theta^{(1)}),\ldots,k(\theta,\theta^{(N)})\right)^T$.

The next step of the BO procedure is the sequential sampling of the input configurations and updating of the parameters of the GP surrogate. Under limited sampling budgets, choosing the next configuration for evaluation $\theta^{N+1}$ is accomplished using the \textit{optimal experimental design}. The goal is to identify regions of the constrained state space $\mathcal{A}$ to draw from regions where the loss function $\mathcal{L}$ is known to have small values (exploitation) or from regions where high uncertainty about the $\mathcal{L}$ values exist (exploration). The next configuration is chosen using a utility function $\mathcal{U}$ which captures the exploration-exploitation trade-off required for the given problem. The expectation over the utility function, known as an acquisition function $a(\theta)$, is taken over the posterior distribution of the GP surrogate function and minimized to obtain the optimal design choice $\theta^{N+1}$ \cite{ryan_contributions_2014}:

\begin{equation}
	\theta^{N+1}=\argmax_{\theta \in \theta_\star} ~ a(\theta) = ~ \mathbb{E}[\mathcal{U}\left(\theta,\Omega)\right]
\end{equation}
where $\theta^{N+1}$ is the optimal input configuration from a pool of potential candidates $\theta_\star$ in the search space $\mathcal{A}$; $\mathcal{U}$ is the chosen utility function, and $\Omega$ represents the hyperparameters of the GP surrogate that approximates the loss function $\mathcal{L}$.	

Several Bayesian utility functions and their non-Bayesian Design of Experiments (DOE) equivalents are discussed extensively in \cite{chaloner_bayesian_1995}. For our calibration problem, we use the Expected Improvement (EI) utility function~\cite{mockus_bayesian_2012}, which weighs the value of a proposed sample by the size of it's potential improvement to avoid getting stuck in local optimas and under-exploration.

\begin{equation}  
\mathcal{U}_{EI}(\theta) = \max [0, \mathcal{L}(\theta^\star) - \mathcal{L}(\theta)]
\end{equation}

The final acquisition function is
\begin{equation}
a_{EI}(\theta) = \Sigma [u\Phi(u) + \phi(u)]
\end{equation}

where $u = \Sigma^{-1}(\mathcal{L}(\theta^\star)-\mu)$, and $\Phi$ is the normal cumulative distribution; $\phi$ is the normal density function, and $\mathcal{N}\left(\mathcal{L};\mu, \Sigma\right)$ is the predictive posterior distribution.

\subsection{Dimension Reduction}\label{sec:DR}
Although GP surrogates approximate non-linear relations between inputs and outputs well, they are known to fail in higher dimensions. Each increase in dimensionality influences the state-space volume exponentially and, in order for Bayesian Optimisation (BO) to maintain reasonable uncertainty in its predictions, this `'curse of dimensionality'' would mandate an exponential growth in sample sizes. Parallelization techniques can expand a limited sampling budget, but only by a multiplicative factor; furthermore, when sample sizes grow significantly, the additional computational complexity within the algorithm itself becomes a hindrance. For this reason, BO is typically ineffective beyond a moderate level of high-dimensional complexity~\cite{sacks_design_1989} when resources are limited. However, if the dimensionality issue can be addressed, BO would become viable again.

Because the relationship between inputs and outputs varies primarily along a small set of curves in many engineering applications, dimensional reduction is a common approach to solve the issue of high-dimensional statespaces. Projection-based techniques find a condensed representation of the dataset in a lower-dimensional subspace and have traditionally been implemented due to their computational and interpretable ease. For linear instances, common techniques include Principal Component Analysis (PCA), Partial Least Squares (PLS), or various single-index models. Of particular popularity in engineering disciplines, the method known as Active Subspaces (AS) projects the data along influential directions known as Principal Components (PCs). Utilizing derivatives, the algorithm identifies the necessary linear combinations to produce a lower-dimensional subspace. For further details of Active Subspaces, see Appendix \ref{Ap:AS}.

As we demonstrated in Section \ref{sec:abmruns}, our ABM exhibits non-linear and significant interactions which would be lost by using these popular linear dimensionality reduction techniques. We propose a non-linear extension of the projection-based methods. Our dimensionality reduction procedures find a map $\phi(\theta) \rightarrow \psi$ that transforms the input vector $\theta$ into a lower-dimensional vector of latent variables $\psi$. We propose using a Deep Learning (DL) architecture to construct the nonlinear mapping $\phi$. Our paper will use Active Subspaces as a benchmark for demonstration.

Deep learners are a class of non-linear functions which construct a predictive mapping using hierarchical layers of latent variables. The output can be a continuous, discrete, or mixed value set. Each layer $\ell$ applies element-wise a univariate activation function $\tau$ to an affine transformation:

\begin{equation}\label{basics}
\begin{split}
Z^{(1)} &=\tau^{(1)}\left(  W^{(0)}\Theta + b^{(0)}\right), \\
\cdots &\\
Z^{(\ell)} &=\tau^{(\ell)}\left(  W^{(\ell -1)}Z^{(\ell -1)} + b^{(\ell - 1)}\right), \\
\hat{Y} & = W^{\ell}Z^{\ell}  + b^{(\ell)},
\end{split}
\end{equation}
where $W$ represents the weights placed on the layer's input set $Z$, and $b$ represents the offset value critical to recovering shifted multivariate functions. Given the number of layers $\ell$, a predictor is a result of the composite map
\begin{equation}\label{Eq:MLP}
\hat{\mathcal{L}} \defeq \left ( \tau_1^{W_1,b_1} \circ \ldots \circ \tau_\ell^{W_\ell,b_\ell} \right ) (\theta).
\end{equation}

The general approximation capabilities of deep learning networks is rooted in the universal basis theorem proved by Kolmogorov~\cite{kolmogorov_approximation_1963}. As data progresses down the network, each layer applies a folding operator to predecessor's divided space according to the chosen activation function $\tau$. Cybenko\cite{cybenko_approximation_1989} proved that a fully-connected network with a single hidden layer and sigmoid activation function can uniformly approximate any continuous multivariate function on a bounded domain within an error margin; further work documented similar results for additional classes of activation functions~\cite{barron_approximation_1994,hornik_approximation_1991,funahashi_approximate_1989} and eventually led to deep learning networks being deemed universal approximators: a single-layered deep learning network with a suitable activation function for continuous functions; a two-layered version for discontinuous and Boolean functions.

Deep learning multi-layered networks designed for dimensionality reduction learning must contain at least one layer $Z^{(i)} \in \mathbb{R}^q$ with fewer nodes than the input $q \ll d$ layer. This bottleneck layer forces the network to learn a compressed, or lower dimensional, set of latent features describing the input-output relationship. We use this representational form to extract a vector representation $\psi$ within one of the deep learner's hidden layers. 

However, to be applicable to our BO framework, our dimensionality reduction technique must be capable of reconstructing a point in the original input space from any point in the latent space. This inversion of the projection map is a non-trivial task for non-linear mappings and is why linear approximation techniques have dominate the field of dimensionality reduction, despite commonly mapping non-linear relationships~\cite{lawrence_gaussian_2004}. We overcome this translation obstacle by incorporating an additional deep learning multi-layered network mapping $s(\psi) \rightarrow \hat{\theta}$ to reconstruct the input vector $\theta$ from the latent variable set $\psi, q \ll d$.

\begin{figure}[H]
	\centering
	\includegraphics[width=0.7\linewidth]{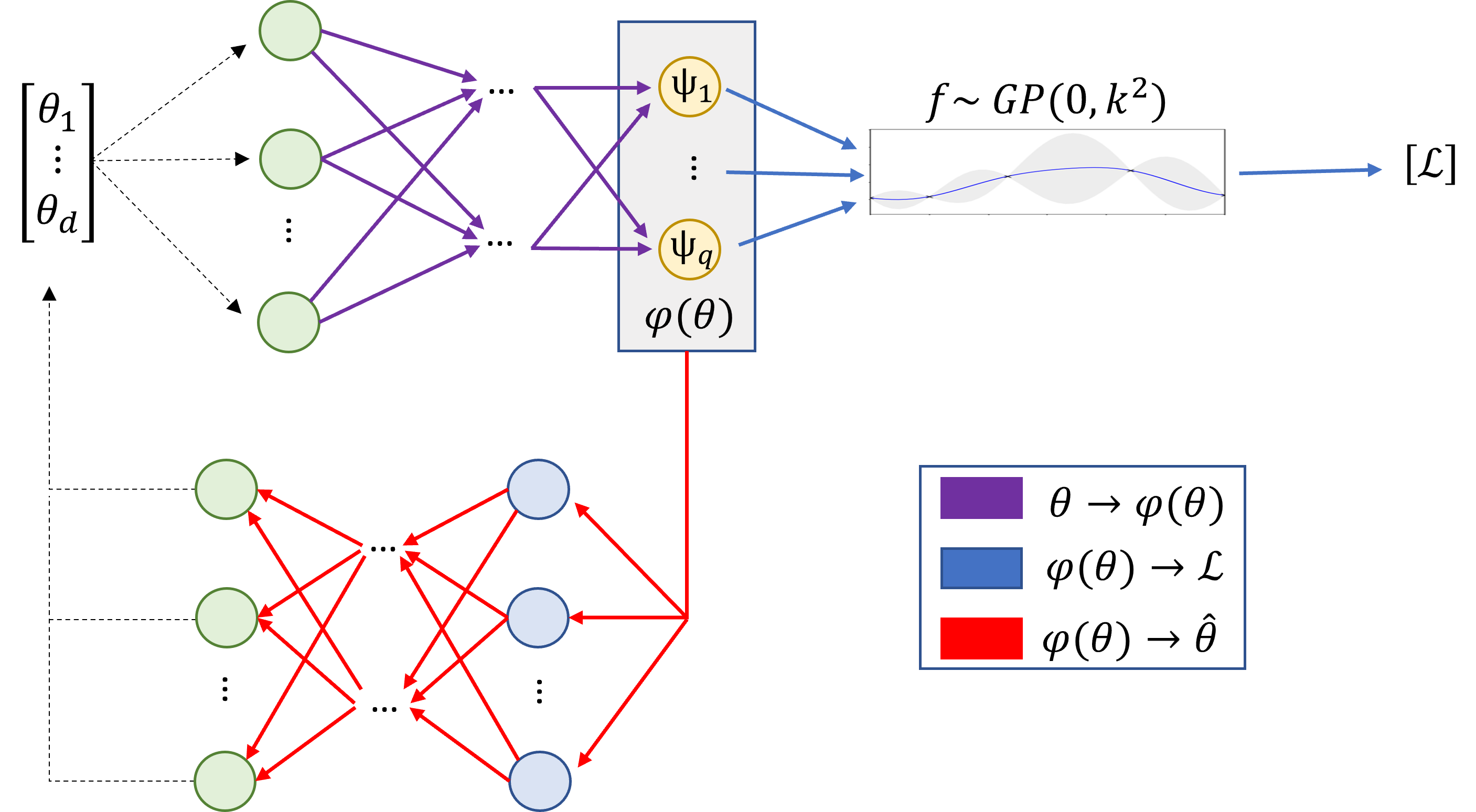}
	\caption{Graphical Representation of the Combinatorial Deep Learning Multi-Layer Network for Calibration}
	\label{fig:MLPAuto}
\end{figure}

The two networks share the same initial layers up to the reduced dimensional layer, as shown in Figure \ref{fig:MLPAuto}, in order to enforce they both use the same encoding substructure. 

\subsection{Combined Model}
The final architecture is then combines dimensionality reduction and GP to approximate the loss function $\mathcal{L}$ and is given by the following equations
\begin{align}
\psi & = \phi_{W_A}(\theta) \label{eq:nn} \\ 
\hat{\theta} & = s_{W_B}(\psi)\ \label{eq:thetahat} \\
\mathcal{L} &\sim \mathcal{GP}(m(\psi), K(\psi,\psi^\prime) \label{eq:gp}.
\end{align}

We define an estimation procedure that jointly estimates the parameters of the deep learning transformation, the basis vectors and hyperparameters of the univariate GPs. Training is accomplished by minimizing a combined training function $\xi(\theta)$ which takes into account both the errors in the predicted output's loss values $\hat{\mathcal{L}}$ and the reconstructed input values $\hat{\theta}$. This joint training of the encoder and decoder allows us to efficiently perform cross validation on the dimensionality of the latent space. 

Though any function which captures these errors can be used, this paper's training function minimizes the Mean Squared Error (MSE) between $\hat{\mathcal{L}}$ and ${\mathcal{L}}$ multiplied by a scaler $\lambda$ and a Mean Squared Error (MSE) function along with a summed quadratic penalty cost $P$ for producing reconstructed values $\hat{\theta}$ outside of the original subspace bounds:

\begin{equation}
\xi(\theta) = - \lambda \sum_{i=1}^N \left( \mathcal{L}_i-\hat{\mathcal{L}}_i  \right)^2
+ \sum_{i=1}^{N}\left[\left(\theta^{(i)}-\hat{\theta}^{(i)}\right)^2 + P\left(\max[0,\hat{\theta^{(i)}}-\theta_{ub}]^2 + \max[0,\theta_{lb}-\hat{\theta^{(i)}}]^2\right)\right],
\end{equation}
 where $\lambda$ represents the importance weight of the encoder verses the decoder portion; $P$ is the penalty cost for producing predicted values outside of the original subspace bounds; $\theta_{ub}$ represents the input set's upper bound; and $\theta_{lb}$ represents the input set's lower bound.

Note, due to the bounded requirement of BO, some activation functions will require adjustments for at least the dimension-reducing bottleneck layer. For example, the ReLU function can be bounded by adjusting the deep learner to add two new parameters defining the state space bounds $DR_{ub},DR_{lb}$:

\begin{equation}
U(\theta^{N+1}) = 
\begin{cases}
DR_{lb}, \ \theta < DR_{lb} \\
\theta, \ DR_{lb} \le \theta \le DR_{lb} \\
DR_{lb}, \ \theta > DR_{lb}
\end{cases}
\end{equation}

When constructing the deep learning architecture, the user should consider that any artificial bounding can result in the need for additional nodes and/or layers when compared with an unbounded version. 

\subsection{Parallelization}

Traditional Bayesian optimization algorithms are sequential in nature~\cite{jones_efficient_1998}; however, when high-dimensional simulation runs take hours or days, maximizing the number of samples become critical to optimization efforts. Several recent approaches address the problem of limited calibration times through parallel evaluations~\cite{wu_parallel_2016,hutchison_parallel_2013,desautels_parallelizing_2014,shah_parallel_2015,wang_parallel_2016} using High Performance Computing (HPC). 

\begin{figure}[H]\vspace*{4pt}
	\centerline{\includegraphics[width=0.5\linewidth]{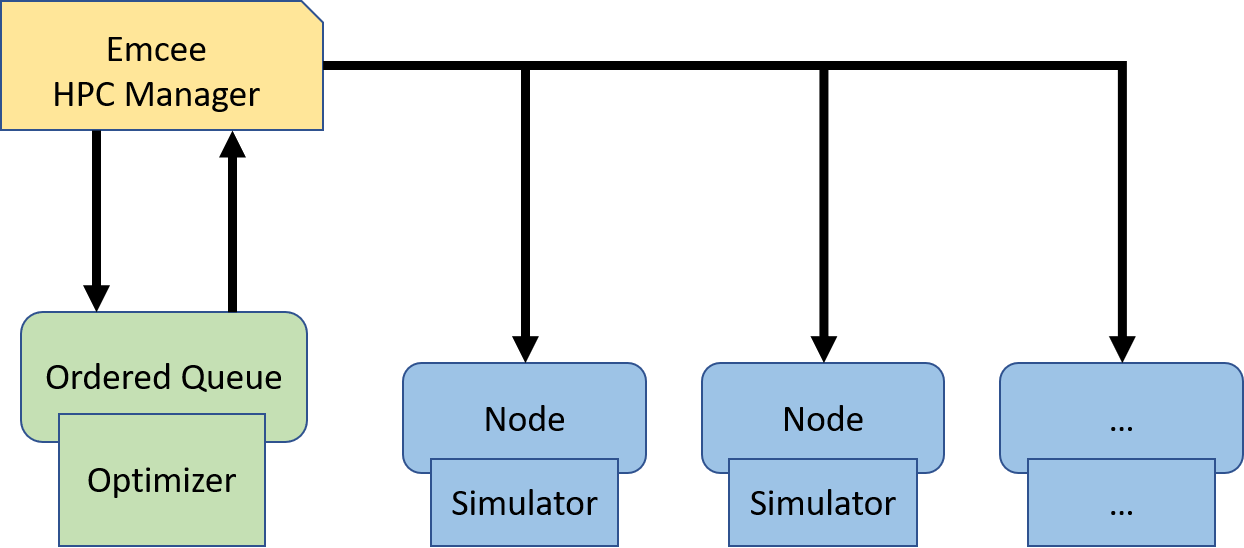}}
	\caption{Computation Framework for HPC}\vspace*{-6pt}
	\label{fig:HPC}
\end{figure}

We incorporate this parallel approach into our computational framework by adding a HPC master controller program to coordinate worker units which run codes across multiple processors concurrently; the result being a larger set of recommended samples within the same limited time frame. Note that, for this optimization framework, the controller must also interact with a model exploration program for its queue of untested input sets, as shown in Figure \ref{fig:HPC}.

In direct consequence of parallelism, BO must also be adjusted to recommend multiple samples. A natural approach may seem to be simply choosing the top $c$ candidates of the acquisition function $a(\theta)$. However, the problem of that approach lies in the assumption that the top set of points depict unique sections of the search space and are not clustered together. This assumption is false. It is not uncommon that, when a new portion of the search space is deemed worth exploring, many of the candidates in that region will contain a large acquisition value, as they all are deemed equally valuable for learning the area's structure.  

Instead, the expected information gained from evaluating a selected candidate within a batch of size $n_{b}$ can be leveraged when deciding the next option. In this approach, the posterior distribution is updated with a pseudo-sample $\hat{\mathcal{L}}$ created from the predicted mean $\mu$ for the chosen candidate $\theta^{(N+1)_i,~i=1,\ldots,n_{b}}$. This expected `'information gain'' now prevents additional candidates from clustering; the posterior distribution is updated before for all subsequent candidates in the batch are chosen.

While proven successful enough to be the most common method implemented\cite{wu_parallel_2016}, this expectation method is not the only possible recourse. Recently, more complex methods have begun to be explored that may prove, in some optimization scenarios where batch sizes are expected to be very large, to be more useful:

\begin{list}{$\bullet$}{}
	\item Gradient applications which choose the next sample by maximizing the expected incremental value~\cite{wu_parallel_2016}, value of information~\cite{xie_bayesian_2016}, or other methods of diversity measurements~\cite{azimi_batch_2012}
	\item Utilizing the lower bound for the previously-chosen but yet-to-evaluate samples~\cite{xie_bayesian_2016}
	\item Partitioning the search space via some design such as treed maximum entropy and selecting candidates from each region proportional to the volume in the region \cite{gramacy_adaptive_2009}
	\item Performing Monte-Carlo simulation to estimate the posterior distribution over examples selected by the sequential method and then select a batch of $c$ examples that “best matches” this distribution; however, no theoretical results prove convergence~\cite{azimi_batch_2012}
\end{list}

\subsection{Adjusted Mean Function}\label{sec:mean}

\begin{figure}\label{Con}
	\centering
 	\includegraphics[width=0.4\linewidth]{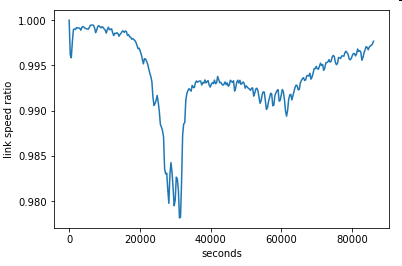}
 	\caption{Traffic Congestion Dependence upon Time}
 	\label{fig:Time}
 \end{figure}

Thus far, the influential relationship between traffic congestion and time, depicted in Figure \ref{fig:Time}, has not been explicitly addressed.

The univariate objective function $\mathcal{L}$ has averaged out the variance across the time dimension and, while the deep learning dimension reduction technique is trained without doing so, the Active Subspace Dimensionality Reduction is not.

We propose leveraging the multivariate capability of deep learning network outputs with the structural influence of a non-zero mean function $m$, not typically used in typical GP applications. Recall in Equation \ref{eq:posterior}, a non-zero mean $m(\theta)$ will not only influence the predictive posterior's mean $\mu$, but also the predictive posterior's covariance $\Sigma$ structure. In addition, because the EI acquisition function $a$ also relies heavily on the predicted mean of a potential candidate, discarding the typical zero-mean assumption will have a critical impact on a candidate's recommendation. Indeed, an educated mean function $\mu_{DL}$ which can provide additional relational information to the distinct application of interest allows for a broadly useful and influential enhancement.

In this paper, we propose using a simple deep learning multi-layered network structure to explicitly capture a deterministic~\cite{polson_deep_2017} mean approximation of the surrogate function in relation to time:

\begin{equation}
    \mathcal{L}(\theta) \sim \mathcal{GP}\left(m_{DL}(\theta),K(\theta,\theta^{\prime})\right)
\end{equation}


\section{Empirical Results}\label{sec:results}

We demonstrate our methodology on a series of simulated examples and then apply it to minimize the ABM calibration objective loss function $\mathcal{L}$, which is the mean squared difference between the observed and simulated turn travel times for every $5$-minute interval across a $24$-hour period.

\subsection{Dimensionality Reduction of Simulated Data}
We begin by demonstrating the quality of our dimensionality reduction approach on two simulated examples (linear and non-linear). Quality of reconstruciton from the reduced space is essential for our calibration algorithm. If unsuccessful, the associated reduction method will likely hinder the optimization algorithm. Therefore, we demonstrate the bi-directional translation of our pre-processing dimension reduction structure on a simple linear and non-linear simulation set with known lower-dimensional structures. Both examples draw on problems from \cite{forrester_engineering_2008}. In this Section, we compare our dimensionality reduction approach with a stat-of-the art technique, called Active Subspace, which is described in an Appendix.

For both examples, we use a set of $n_{train}=80$ training samples for learning the encoders and decoders and $n_{test}=200$ testing samples to assess quality of reconstruction. Both training and testing samples are generated using Latin Hypercube sampling. For the Active Subspace technique, we use the author's recommended local linear approximation documented in \cite{constantine_active_2015}, which uses least-squares to fit the coefficients $\hat{\beta_0}$ and $\hat{\beta}$ of the model: 
\[
	\mathcal{L}(\theta)\approx {\hat{\beta}}_{0}+\hat{\beta}^{T}\theta. 
\]
Then, the gradient required for Active Substpace is approximated as $\nabla_\theta \mathcal{L}(\theta)= \hat{\beta}$.
In addition, $500$ bootstraps are used for each example and we execute the radial basis response surface code from the \href{https://github.com/paulcon/active_subspaces}{Active Subspaces Python package} produced by the method authors to map the relationship between reduced input vector $\psi_{AS}$ and the true function output $y$. Lastly, to provide benchmark comparisons, and because Active Subspaces is the only method which requires self-determination of the dimensional size of its reduced subspace, this paper uses the final Active Subspace dimension as the limit $k_{max}$ which the deep learning model can project up to. Each example will note the dimensions.

\subsubsection{Linear Function}
	
The first example has linear simulated response $y$, a $1$-dimensional output, given a $10$-dimensional input set $\theta=[\theta_1,\cdots,\theta_{10}]$ between $[0,1]$:
	
	\begin{equation}
	\begin{split}
		\mathcal{L}(\theta) &= \sum_{i=1}^{4}0.2\theta_{i} + 0\theta_5 + 0\theta_6 + 0\theta_7 + 0\theta_8 + 0\theta_9 + 0\theta_{10} + e\\
	e &\sim \mathcal{N}(0,0.05^2)
	\end{split}
	\end{equation}

\begin{figure}[H]
		\centering
		\begin{tabular}{ccc}
			\includegraphics[width=0.3\linewidth]{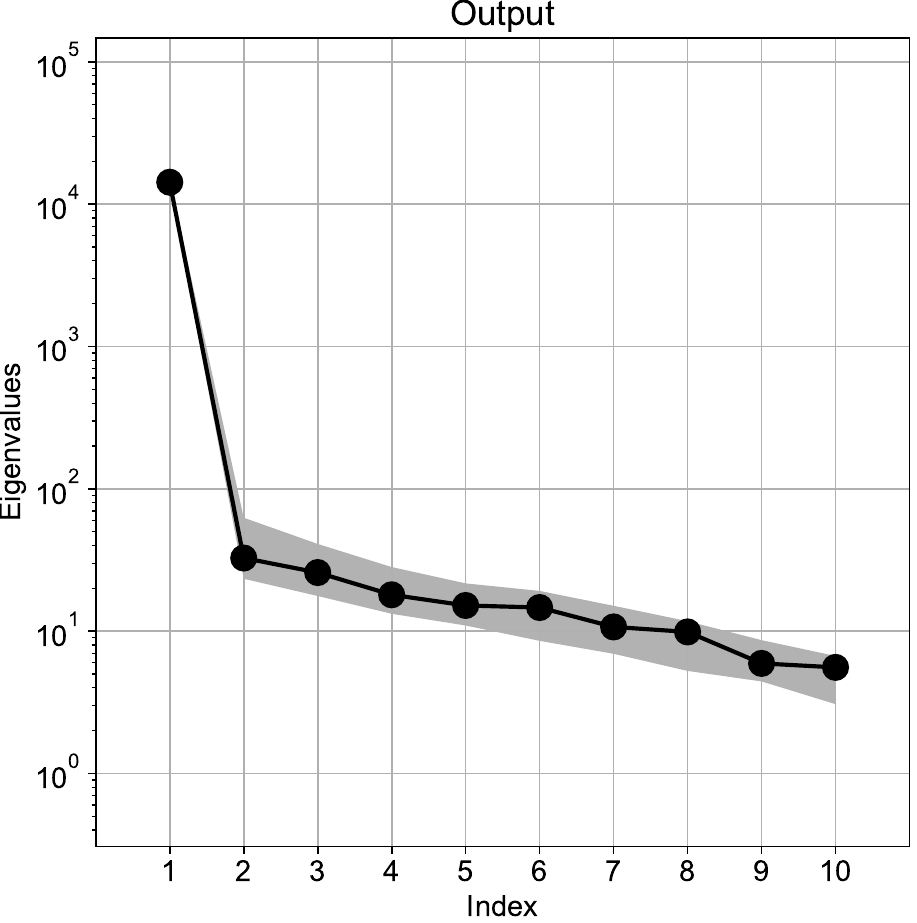} &
			\includegraphics[width=0.3\linewidth]{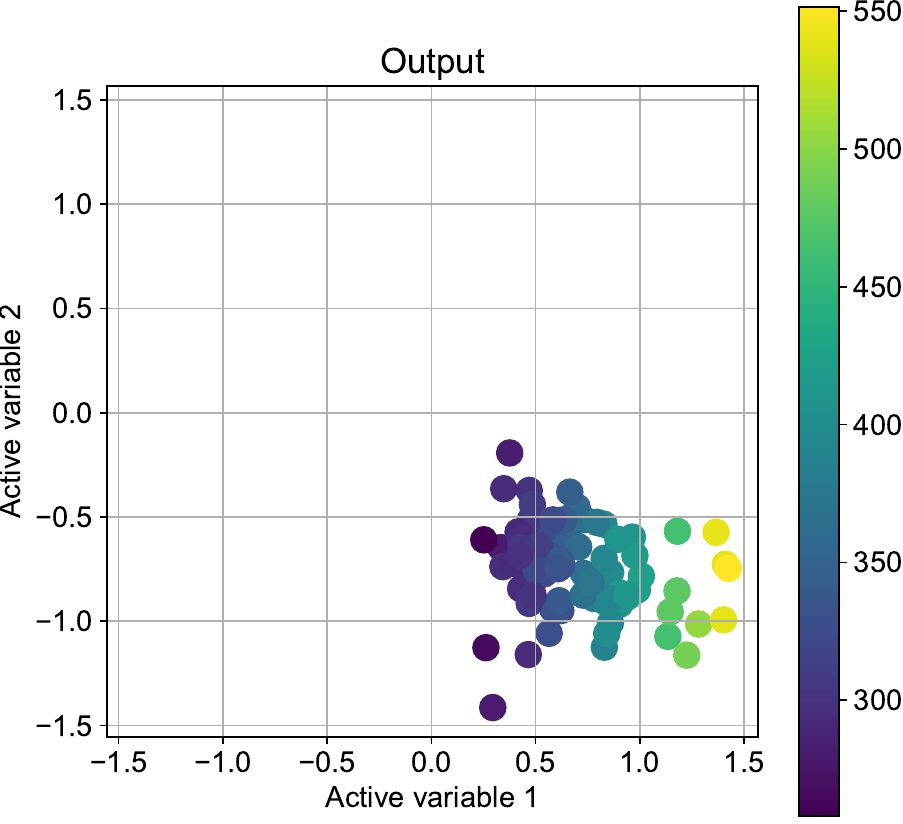} &
			\includegraphics[width=0.3\linewidth]{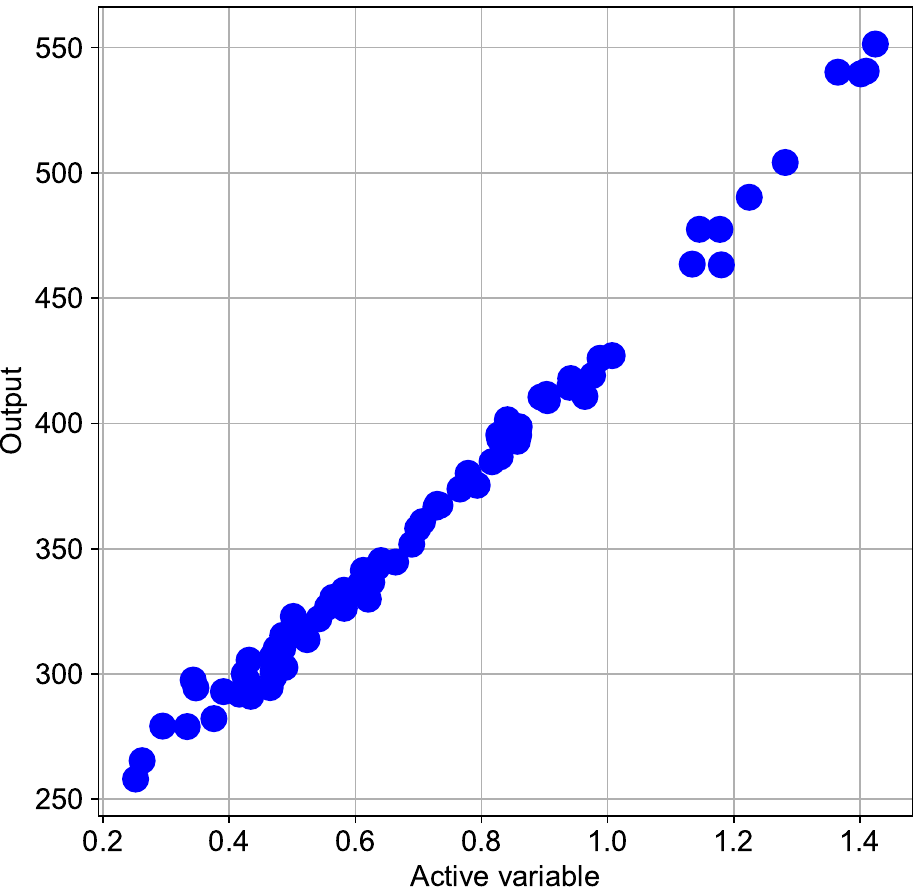}\\
			(a) Eigenvalues & (b) 1-D Active Subspace & (c) 2-D Active Subspace
		\end{tabular}
		\caption{(a) visualizes the eigenvalues of each eigenvector; the significant separation between the first and second index indicates a single dimensional active subspace (b) illustrates the output trend over a varying 1-D active subspace; the linear trend signifies a 1-D subspace is successful (c) illustrates the output trend over a varying 2-D active subspace; the output shows to only be influenced by the first active subspace regardless of the second's change in values.}
		\label{fig:Ex1_AS_gap}
	\end{figure}

In this manner, a lower-dimensional representation of the input data is guaranteed to be producible. The Active Subspace method $AS$ found a significant 'gap' between the first and second subspace dimensions, as shown in Figure \ref{fig:Ex1_AS_gap}(a), and, therefore, determined its dimension set to be $k=1$; the linear trend in Figure \ref{fig:Ex1_AS_gap}(b) and the second dimension's lack of influence over the first, displayed in Figure \ref{fig:Ex1_AS_gap}(c), corroborates this conclusion. As a result, our proposed deep learning model must also use $k=1$ dimensions for this problem.
	
For the deep learning technique $DL$, a simple multi-layered network trained on $1000$ epochs with learning rate $\alpha=0.01$ and seed=$13$ is used with the following structure:
	
	\begin{figure}[H]
		\centering
		\includegraphics[width=0.50\linewidth]{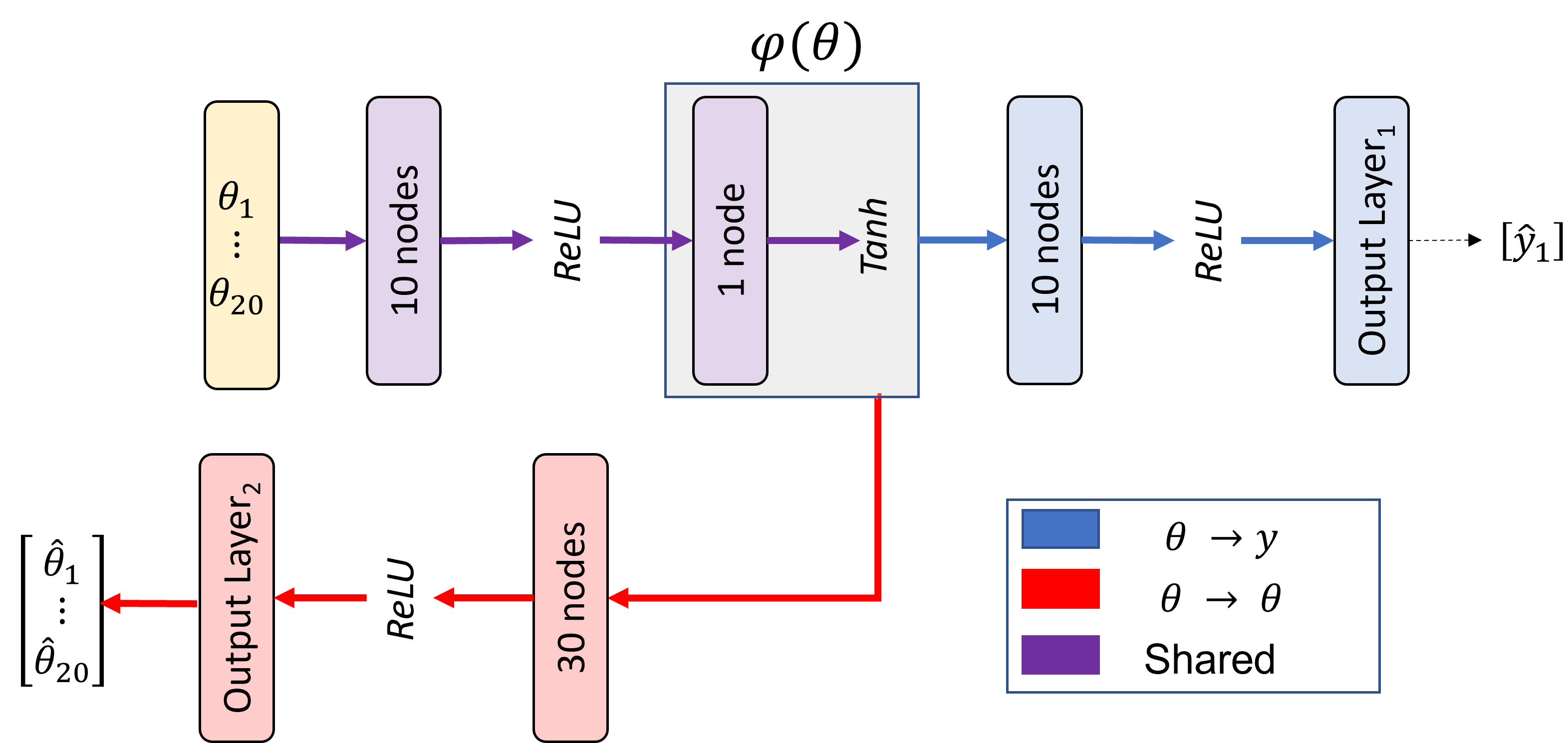}
		\caption{Example 1 Deep Learning Multi-layered Network Structure}
	\end{figure}
	
We set $\lambda = 100,~ P = 100$ to make decoding and encoding equally important. 

The authors would like to note that the same results can be found using a purely linear deep learner with a learning rate of $\alpha=0.001$:
	
	\begin{figure}[H]
		\centering
		\includegraphics[width=0.40\linewidth]{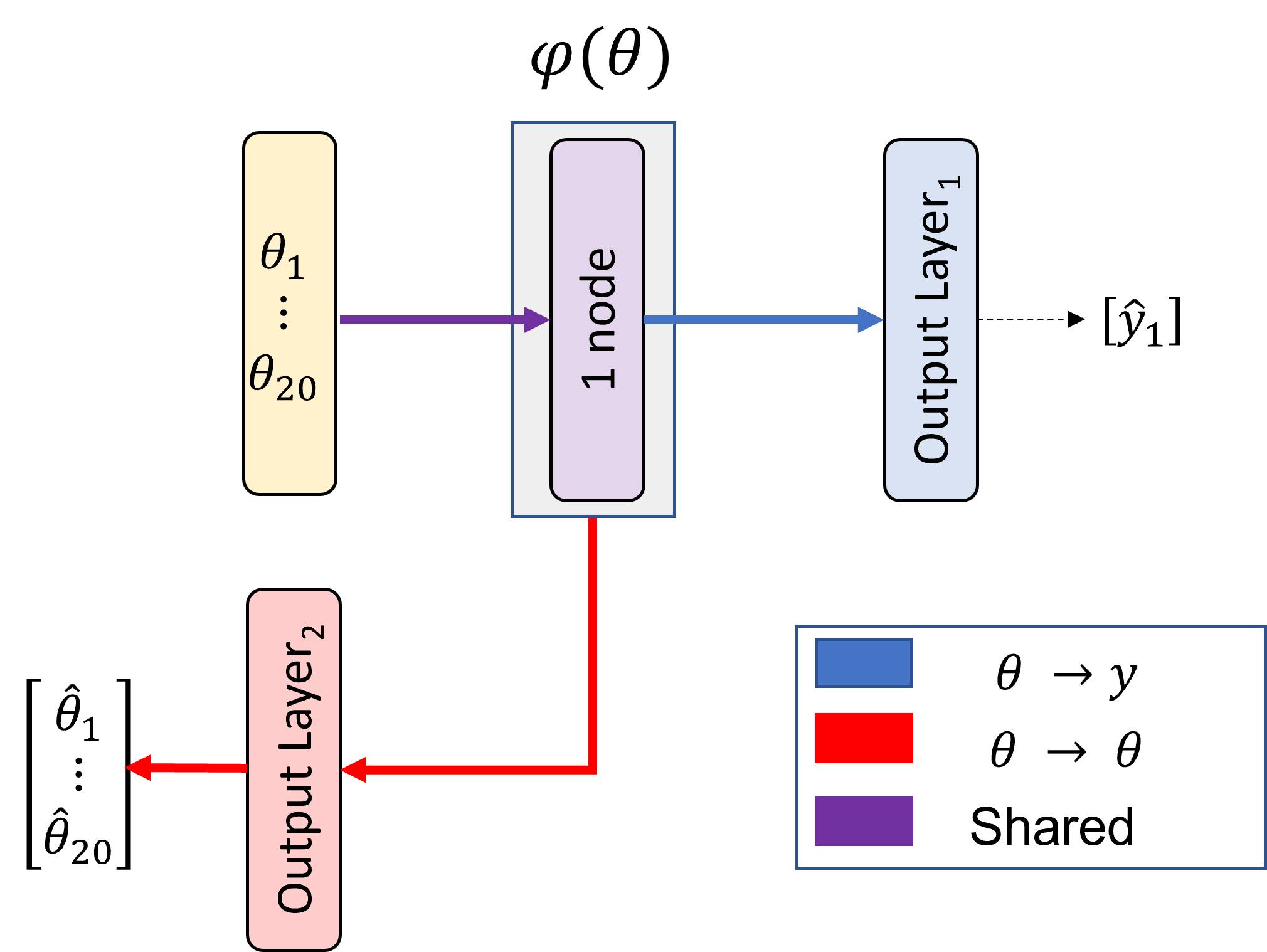}
		\caption{Alternative Example 1 Deep Learner Multi-layered Network Structure}
	\end{figure}
	
The non-linear version is used, however, to demonstrate and emphasize that a nonlinear neural network is still capable of capturing a linear relationship.

\begin{figure}[H]
	\centering
	\begin{tabular}{cc}
		\includegraphics[width=0.3\linewidth]{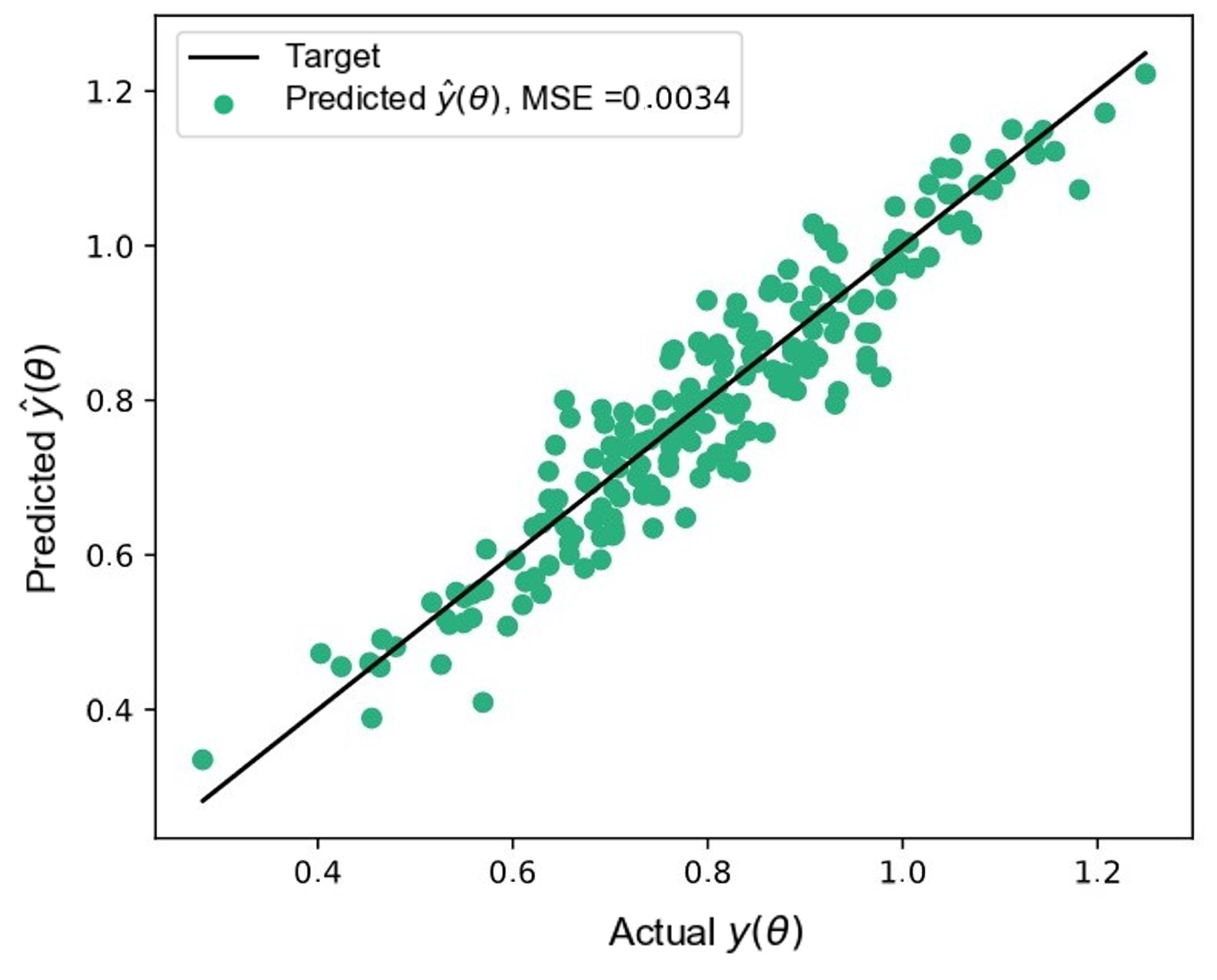} &
		\includegraphics[width=0.3\linewidth]{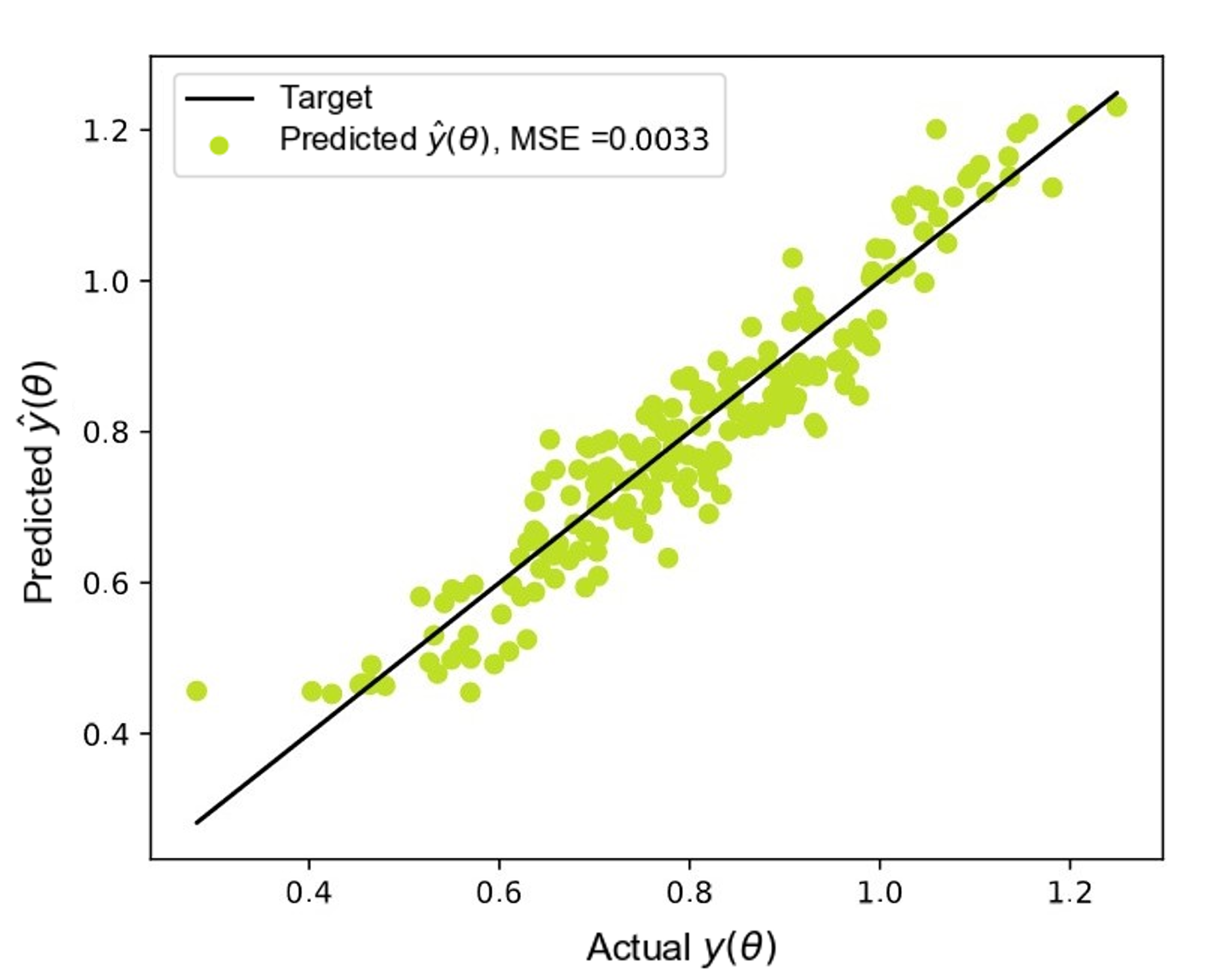} \\
		(a) Active Subspace & (b) Deep Learner
	\end{tabular}
	\caption{Comparing the accuracy of the (a) $AS$ dimension reduction architecture and (b) $DL$ dimension reduction architecture's predictions for $y$. The Mean Squared Error (MSE) provides benchmark comparisons while the black line indicates the correct position if $\hat{y}(\theta) = y(\theta)$.}
	\label{fig:Ex1_a}
\end{figure}

To begin, we look at how well the predicted outputs $\hat{y}$ align with the true outputs $y$ for each method. Shown in Figure \ref{fig:Ex1_a}, the two methods provide similar results with a MSE of $0.0033$ for the $AS$ method and $0.0034$ for the $DL$ method. However, the $DL$ method produces stronger predictions about the center of the statespace, where most of the samples concentrate, while $AS$ produces more accurate predictions at either end. This behavior is likely reflective of how the two subspaces are found: $AS$ seeks the directions of average variability across the entire statespace; $DL$ wants to minimize its error among the bulk of samples.

Having set the importance of learning the encoding equal to decoding while training, we would expect similar success when analyzing the accuracy of the methods' decoder. To check, we first decode the encoded lower-dimensional representations $\psi$ into estimated reconstructions $\hat{\theta}$ and then evaluate them through the true function, $Y(\hat{\theta})$. We take this approach because multiple samples can produce the same output, making their lower representation indistinguishable. For this reason, the decoded input is unlikely to match the original inputs exactly, but should be expected to produce the same outcomes when run through the true function.

\begin{figure}[H]
	\centering
	\begin{tabular}{cc}
		\includegraphics[width=0.3\linewidth]{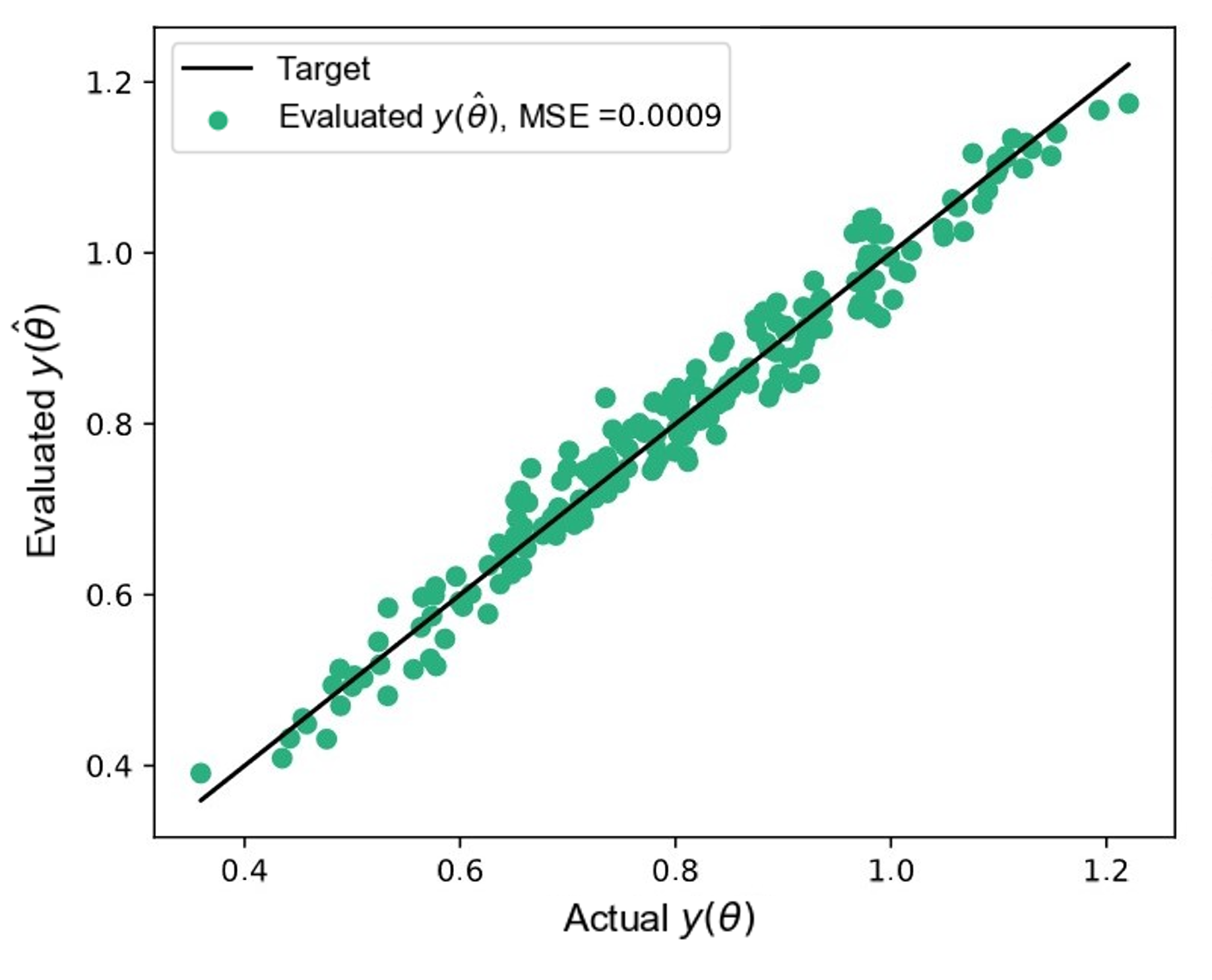} &
		\includegraphics[width=0.3\linewidth]{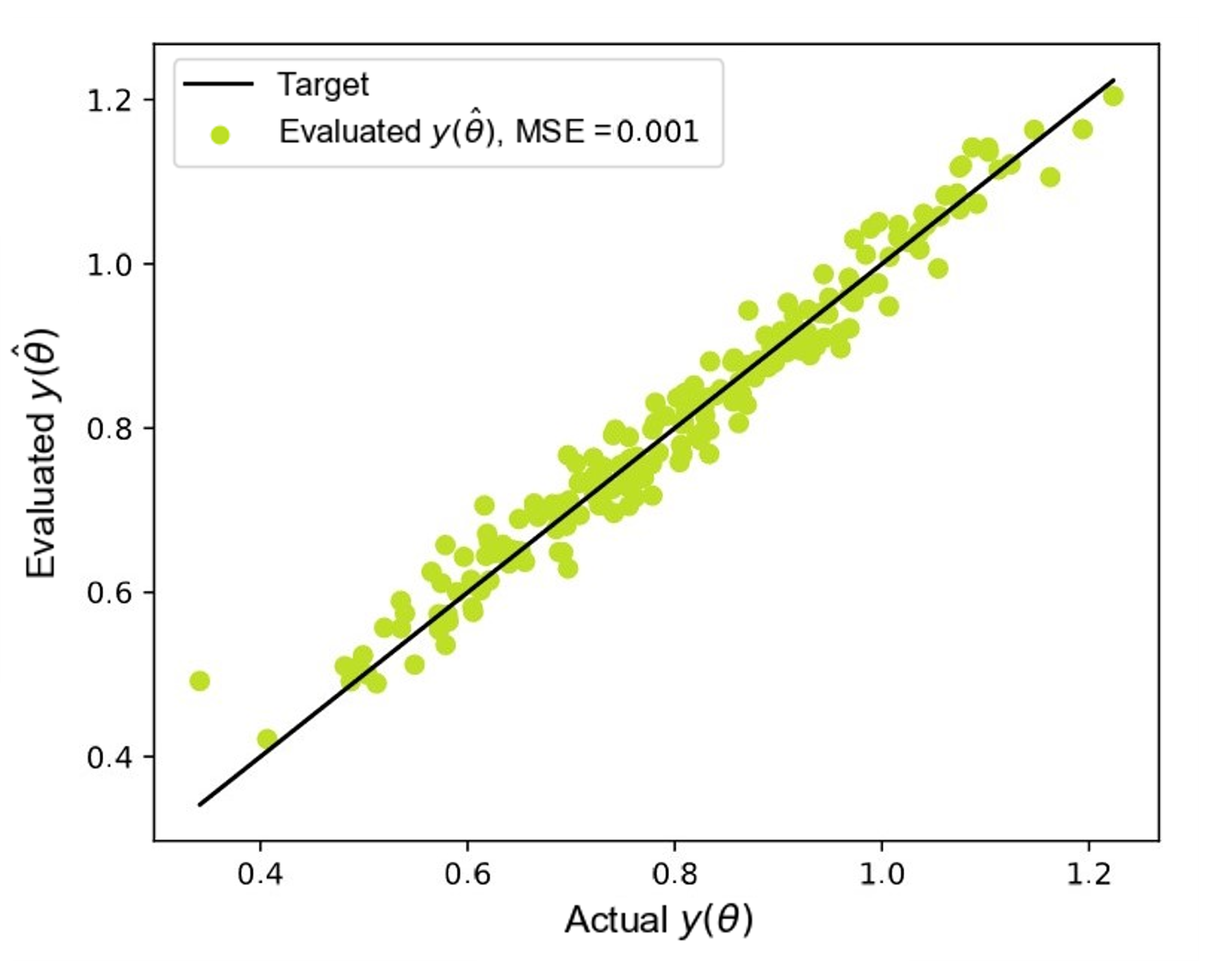} \\
		(a) Active Subspace & (b) Deep Learner
	\end{tabular}
	\caption{Comparing the success of the (a) $AS$ dimension reduction decoder and (b) $DL$ dimension reduction decoder in capturing non-linearity by comparing the simulated outputs given the decoded inputs $y(\hat{\theta})$ and the simulated output given the original inputs $y(\theta)$.The Mean Squared Error (MSE) provides benchmark comparisons while the black line indicates the correct position if $y(\hat{\theta}) = y(\theta)$ and capturing non-linearity was successful.}
	\label{fig:Ex1_b}
\end{figure}

Note, the true $\theta$ values are simultaneously re-evaluated in the simulator with the reconstructed $\hat{\theta}$ to ensure they share the exact random error; this way, a perfect match would be reflected at $Y{\hat{\theta}}=Y{\theta}$. The results are shown in Figure \ref{fig:Ex1_b} and, as expected, similar success is found with an MSE of $0.0009$ for the $AS$ method and $0.001$ for the $DL$ method.

\begin{figure}[H]
	\centering
	\begin{tabular}{ccc}
		\includegraphics[width=0.3\linewidth]{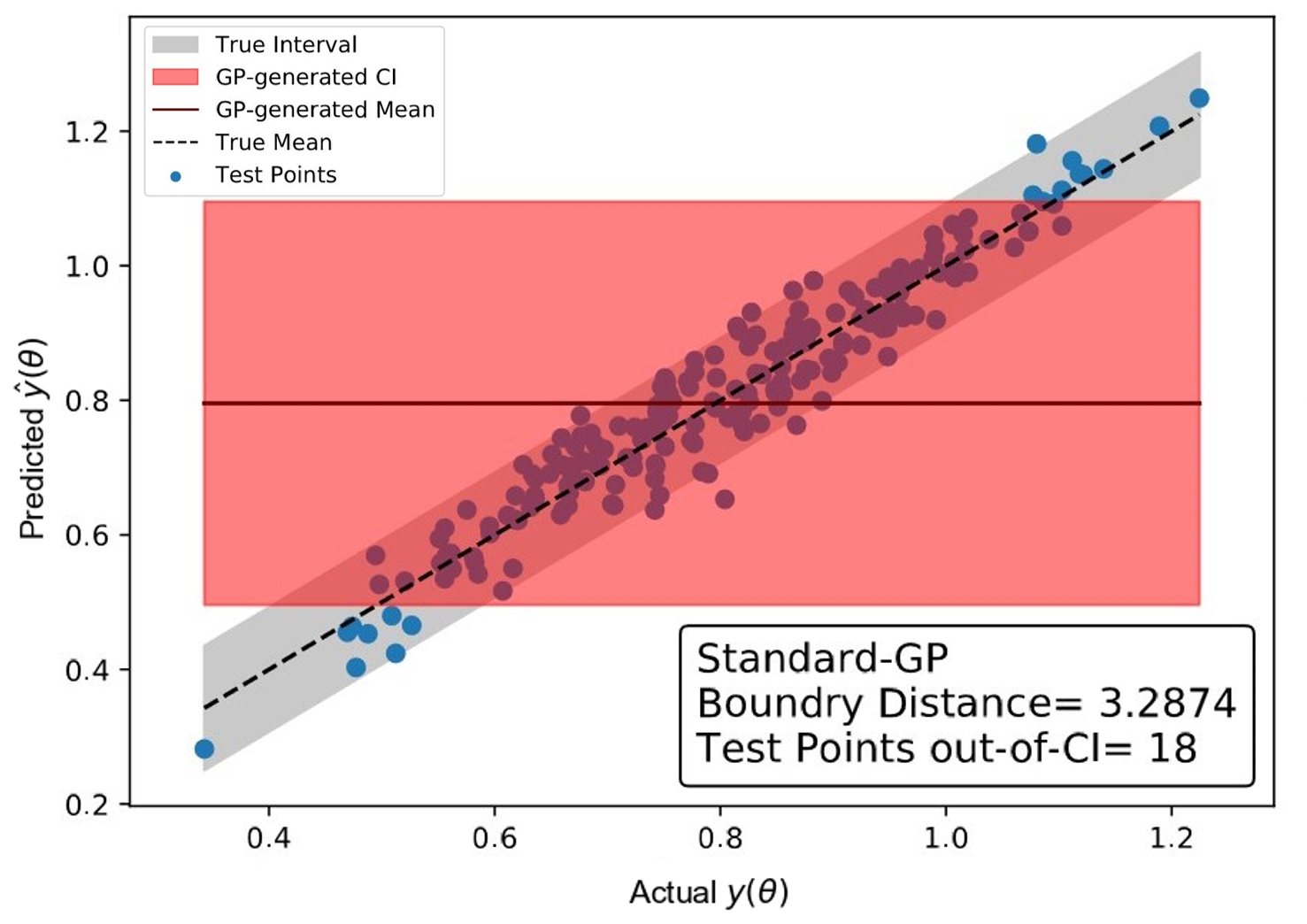} &
		\includegraphics[width=0.3\linewidth]{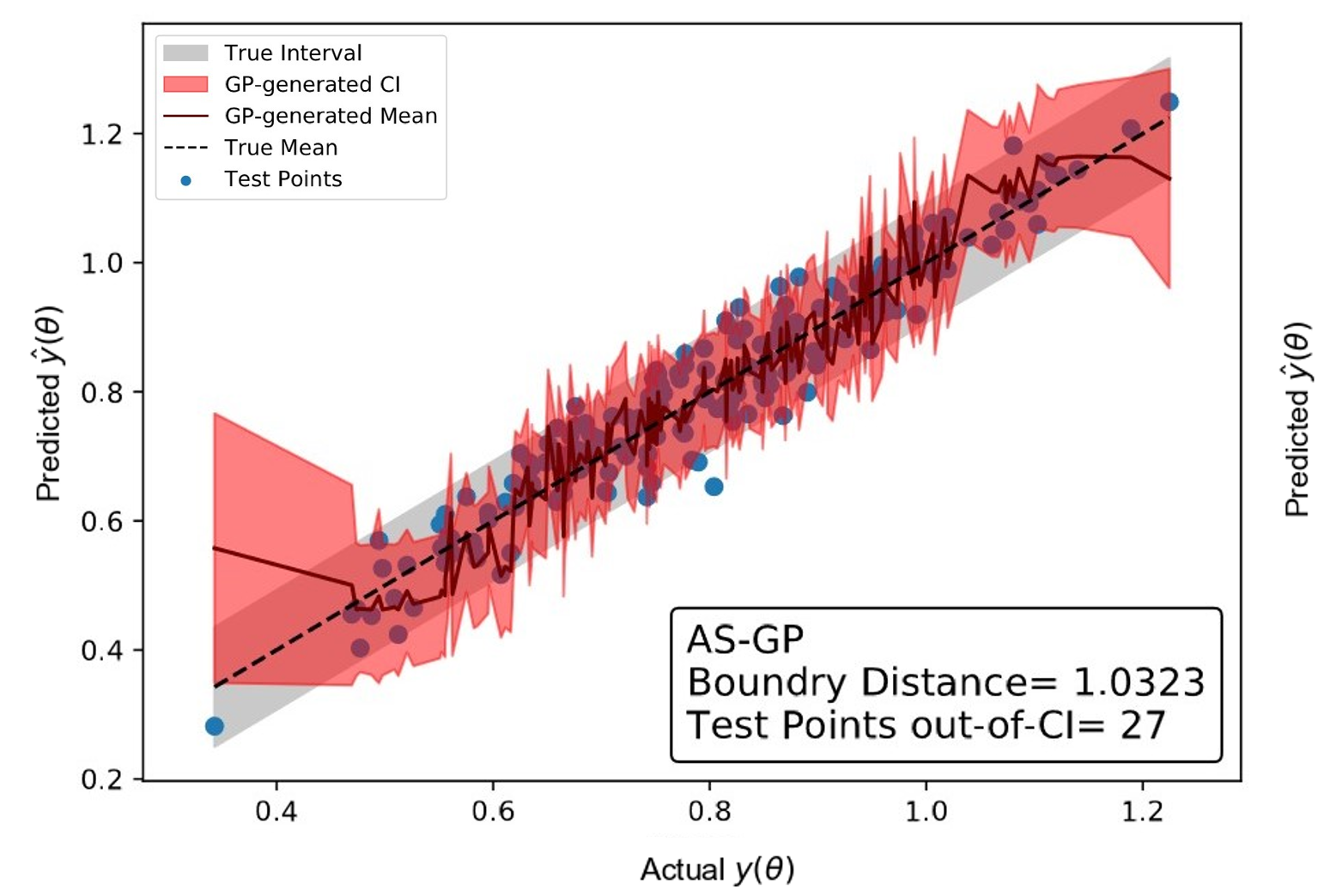} &
		\includegraphics[width=0.3\linewidth]{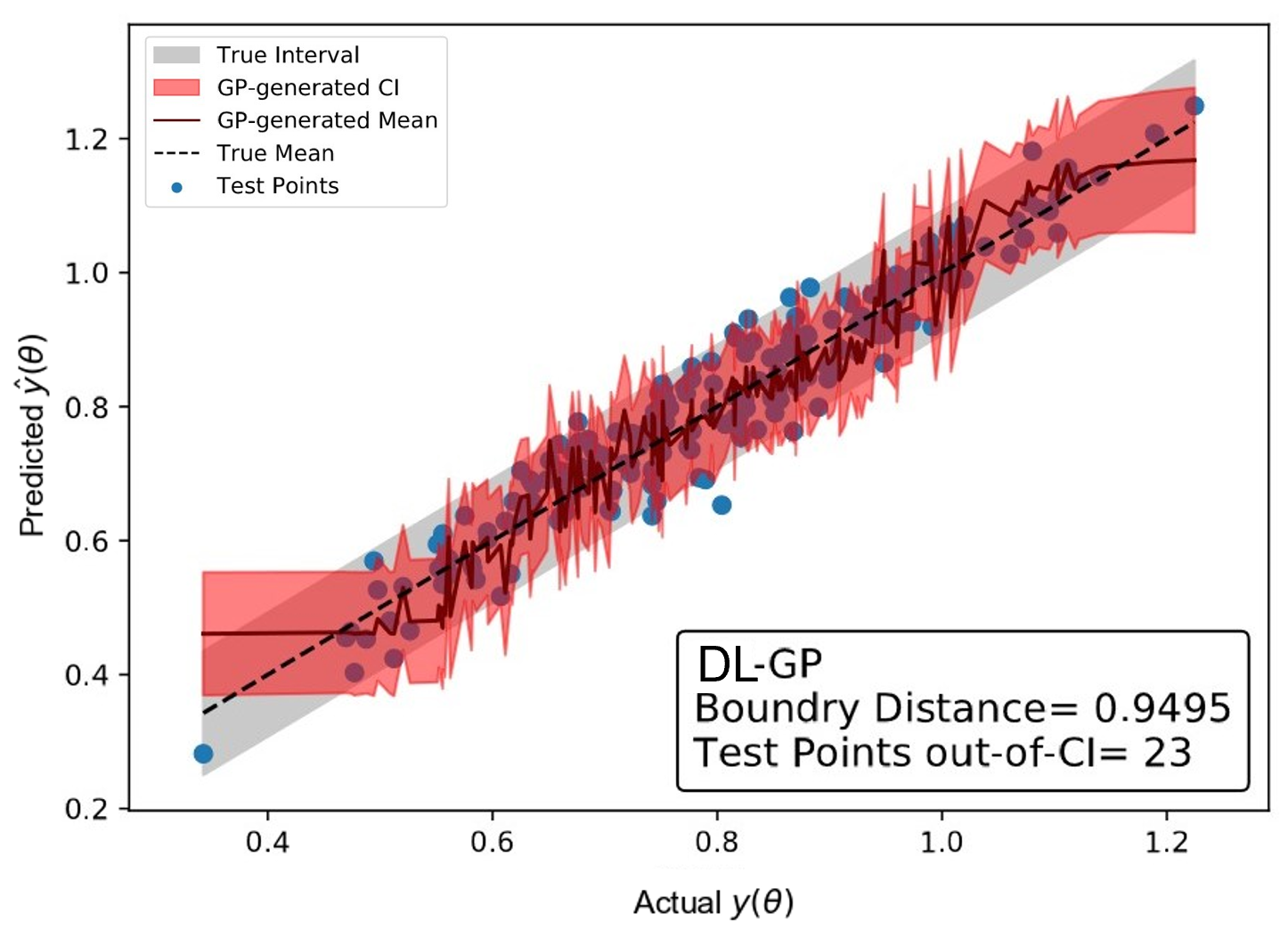} \\
		(a) $Original$ + $\mathcal{GP}_{0}$ & (b) $AS$ + $\mathcal{GP}_{0}$ & (c) $DL$ + $\mathcal{GP}_{0}$
	\end{tabular}
	\caption{Comparison of $n_{test}=200$ test predictions by a constructed GP based on each reduction method with the true values. Since the true function is stochastic, the true values are captured as follows: the black line indicates the correct position if the predicted value mean matches the true mean; the gray shaded region indicates the true function's $95\%$ Confidence Interval; and the blue points are the evaluations provided as test data. The red line represents the mean predicted value of the GP; the red shaded region indicates the GP's $95\%$ Confidence Interval.The GP is built based on only the training set $n_{train}=80$.}
	\label{fig:Ex1_c}
\end{figure}

Finally, Figure \ref{fig:Ex1_c} shows what a zero-mean GP built over the two reduced subspaces, referred as $AS$ + $\mathcal{GP}_{0}$ and $DL$+$\mathcal{GP}_{0}$, would predict for our $200$ test data points and further compare them with a zero-mean GP built over the original subspace $Original$ + $\mathcal{GP}_{0}$, i.e. without any dimension reduction pre-processing. To provide quantified context across the three GPs instances, the authors offer both the number of points which lie outside of the predicted $95\%$ Confidence Intervals graphed in red, and a 'distance' value describing the magnitude difference of the various predictive Confidence Intervals against the true one. 
\begin{equation}
\mathrm{BoundaryDistance}_{DR}= \sum_{i=0}^{200} \frac{\sqrt{(\hat{LB}_{DR}(\theta^{(i)}) - y^{(i)})^2} + \sqrt{(\hat{UB}_{DR}(\theta^{(i)}) - y^{(i)})^2}}{	\sqrt{(LB(\theta^{(i)}) - y^{(i)})^2} + \sqrt{(UB(\theta^{(i)})-y^{(i)})^2
}}
\end{equation}

where $\hat{LB},\hat{UB}$ are the estimated bounds of the predicted confidence interval produced by the GP built on dimension reduction technique $DR$; $LB,UB$ are the true bounds based on the problem, and $\theta^{(i)},y^{(i)}$ are the test data sets.

Most notably, both pre-processing methods achieve dramatic improvements to the GP built using the original $10$-dimensional state-space $Original$ + $\mathcal{GP}_0$ , highlighting the advantage of our projection-based approach. Despite its more accurate predictions $\hat{y}$ at the tail ends of the statespace, the $AS$ + $\mathcal{GP}_0$ build is unexpectedly less confident;and, its predicted mean veers more significantly from true average than the $DL$ + $\mathcal{GP}_0$. Due to the volume of samples clustered in the middle, this divergence is only hinted at by the slight difference in the distance calculation of $1.0323$ for $AS$ + $\mathcal{GP}_0$ and $0.9495$ for $DL$ + $\mathcal{GP}_0$.

	
	
\subsubsection{Non-linear Prediction}\label{sec:NL}
	
The second example produces a non-linear simulated response $y$, a $1$-dimensional output, from a $10$-dimensional input set $\theta=[\theta_1,\cdots,\theta_{10}]$:	\begin{equation}
\begin{split}
	\mathcal{L}(\theta) &= \sin{(\theta_{1})} + \sin{(5\theta_{2})} + 0\theta_3 + 0\theta_4 + 0\theta_5 + 0\theta_6 + 0\theta_7 + 0\theta_8 + 0\theta_9 + 0\theta_{10} + e \\
	e &\sim \mathcal{N}(0,0.05^2)
	\end{split}
	\end{equation}

\begin{figure}[H]
		\centering
		\begin{tabular}{ccc}
			\includegraphics[width=0.3\linewidth]{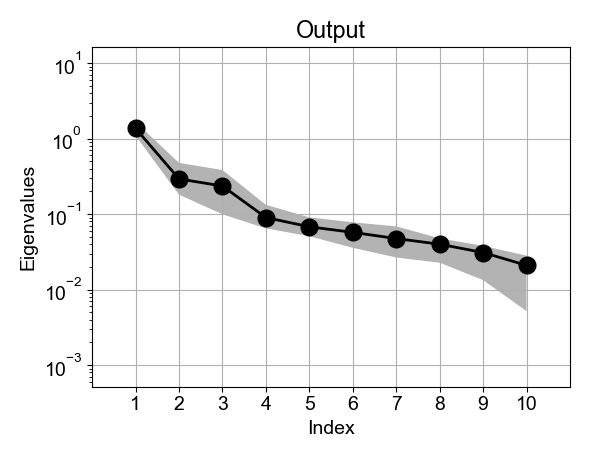} &
			\includegraphics[width=0.3\linewidth]{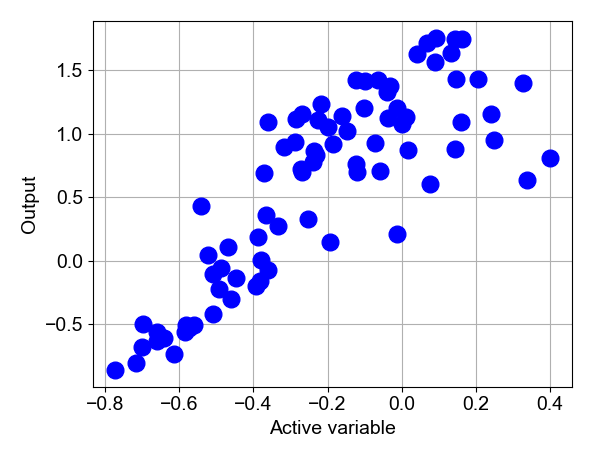} &
			\includegraphics[width=0.3\linewidth]{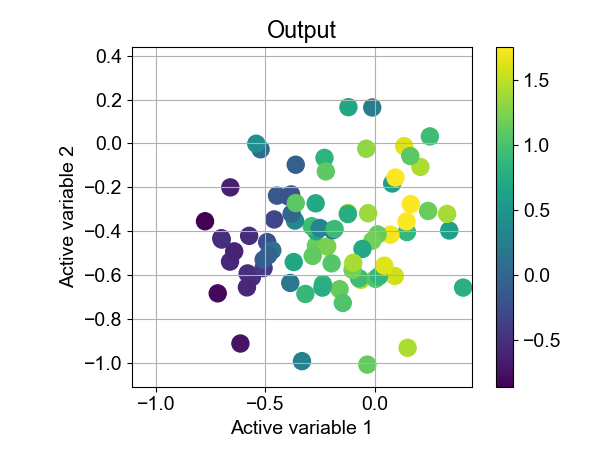}\\
			(a) Eigenvalues & (b) 1-D Active Subspace & (c) 2-D Active Subspace
		\end{tabular}
		\caption{(a) visualizes the eigenvalues of each eigenvector; the significant separation between the first and second index indicates a single dimensional active subspace. However, the (b) 1-D active subspace illustrates the output trend is not linear and that the second index may be necessary. (c) illustrates the output trend over a varying 2-D active subspace; the output supports the use of two active subspaces.}
		\label{fig:Ex2_AS_gap}
	\end{figure}

The Active Subspace method determined a dimension set of $k=2$, as shown in Figure \ref{fig:Ex2_AS_gap}. 
	
Our deep learner uses a learning rate $\alpha=0.01$ and $seed=88$ under the following structure:
	
	\begin{figure}[H]
		\centering
		\includegraphics[width=0.50\linewidth]{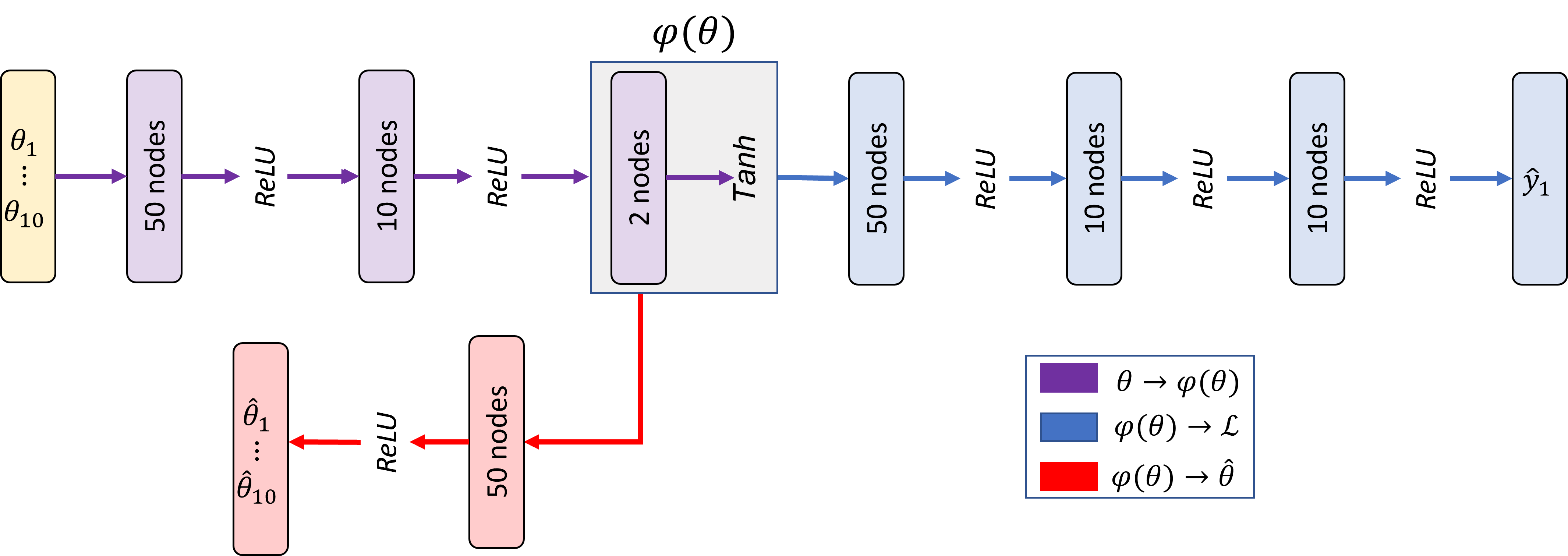}
		\caption{Example 2 Neural Network Structure}
	\end{figure}
	
By setting $\lambda = 10$, we again make decoding and encoding equally important.
	
\begin{figure}[H]
	\centering
	\begin{tabular}{cc}
		\includegraphics[width=0.3\linewidth]{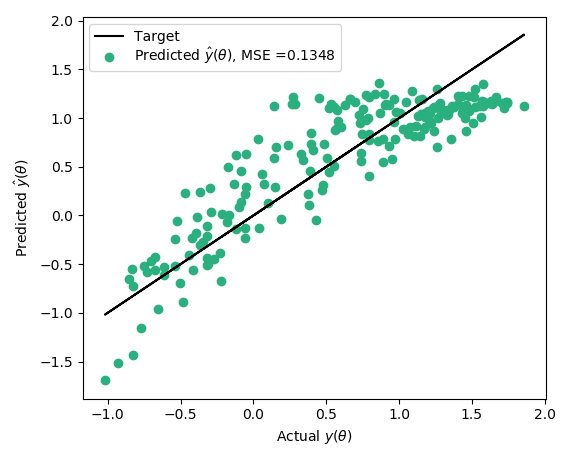} &
		\includegraphics[width=0.3\linewidth]{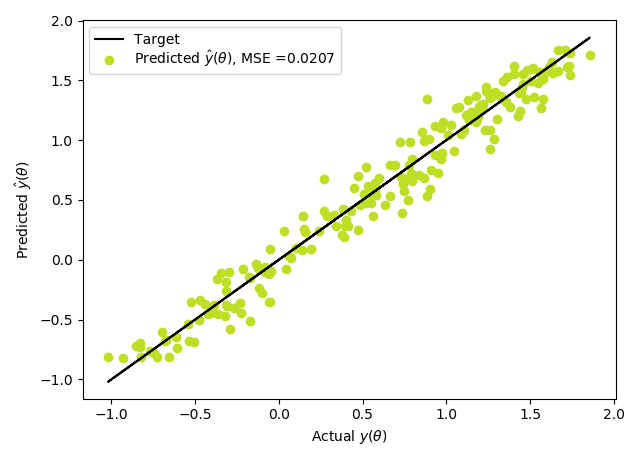} \\
		(a) Active Subspace & (b) Deep Learner
	\end{tabular}
	\caption{Comparing the accuracy of the (a) $AS$ dimension reduction architecture and (b) $DL$ dimension reduction architecture's predictions for $y$. The Mean Squared Error (MSE) provides benchmark comparisons while the black line indicates the correct position if $\hat{y}(\theta) = y(\theta)$.}
	\label{fig:Ex2_a}
\end{figure}

Figure \ref{fig:Ex2_a} displays the results of predicting the output $\hat{y}$ from each method's lower-dimensional transformation $psi$.
	
The curved effect seen in the Figure \ref{fig:Ex2_a}(a) is the consequence of the $AS$ technique's difficulty with capturing the non-linear behavior of the problem; the $DL$ method, on the other hand, performs exceedingly well, producing predictions clustered about the true values for the entire domain.

\begin{figure}[H]
	\centering
	\begin{tabular}{cc}
		\includegraphics[width=0.3\linewidth]{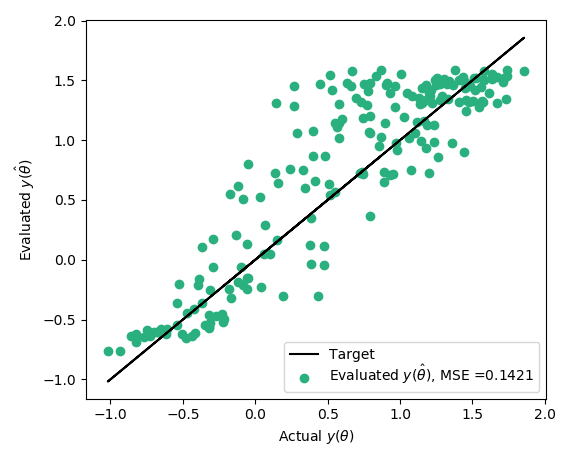} &
		\includegraphics[width=0.3\linewidth]{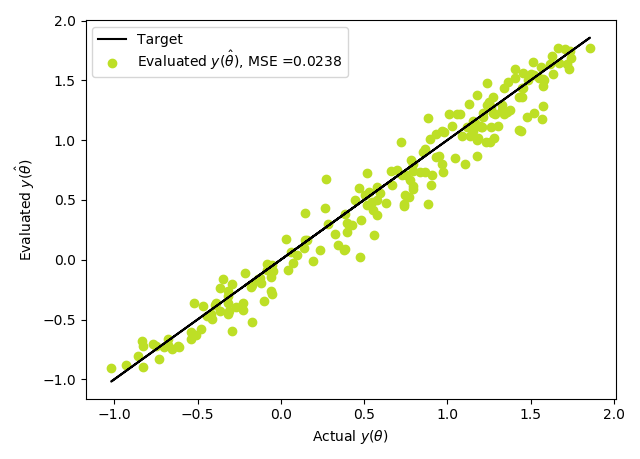} \\
		(a) Active Subspace & (b) Deep Learner
	\end{tabular}
	\caption{Comparing the success of the (a) $AS$ dimension reduction decoder and (b) $DL$ dimension reduction decoder in capturing non-linearity by comparing the simulated outputs given the decoded inputs $y(\hat{\theta})$ and the simulated output given the original inputs $y(\theta)$.The Mean Squared Error (MSE) provides benchmark comparisons while the black line indicates the correct position if $y(\hat{\theta}) = y(\theta)$ and capturing non-linearity was successful.}
	\label{fig:Ex2_b}
\end{figure}

Figure \ref{fig:Ex2_b} displays the results of the simulated outcomes using each method's decoded input set $y(\hat{\theta})$. The $DL$ method remains strong in its reconstruction capabilities, with a similar MSE of $0.02$ while the $AS$ method continues to have difficulty with the problem, though the more obvious curvature of the predictive outputs is absent.

Lastly, we repeat the previous example's final analysis via Figure \ref{fig:Ex1_c} to show what a zero-mean GP built over the two reduced subspaces, referred as $AS$ + $\mathcal{GP}_{0}$ and $DL$+$\mathcal{GP}_{0}$, would predict for our $200$ test data points and further compare them with a zero-mean GP built over the original subspace $Original$ + $\mathcal{GP}_{0}$, i.e. without any dimension reduction pre-processing. 

\begin{figure}[H]
	\centering
	\begin{tabular}{ccc}
		\includegraphics[width=0.3\linewidth]{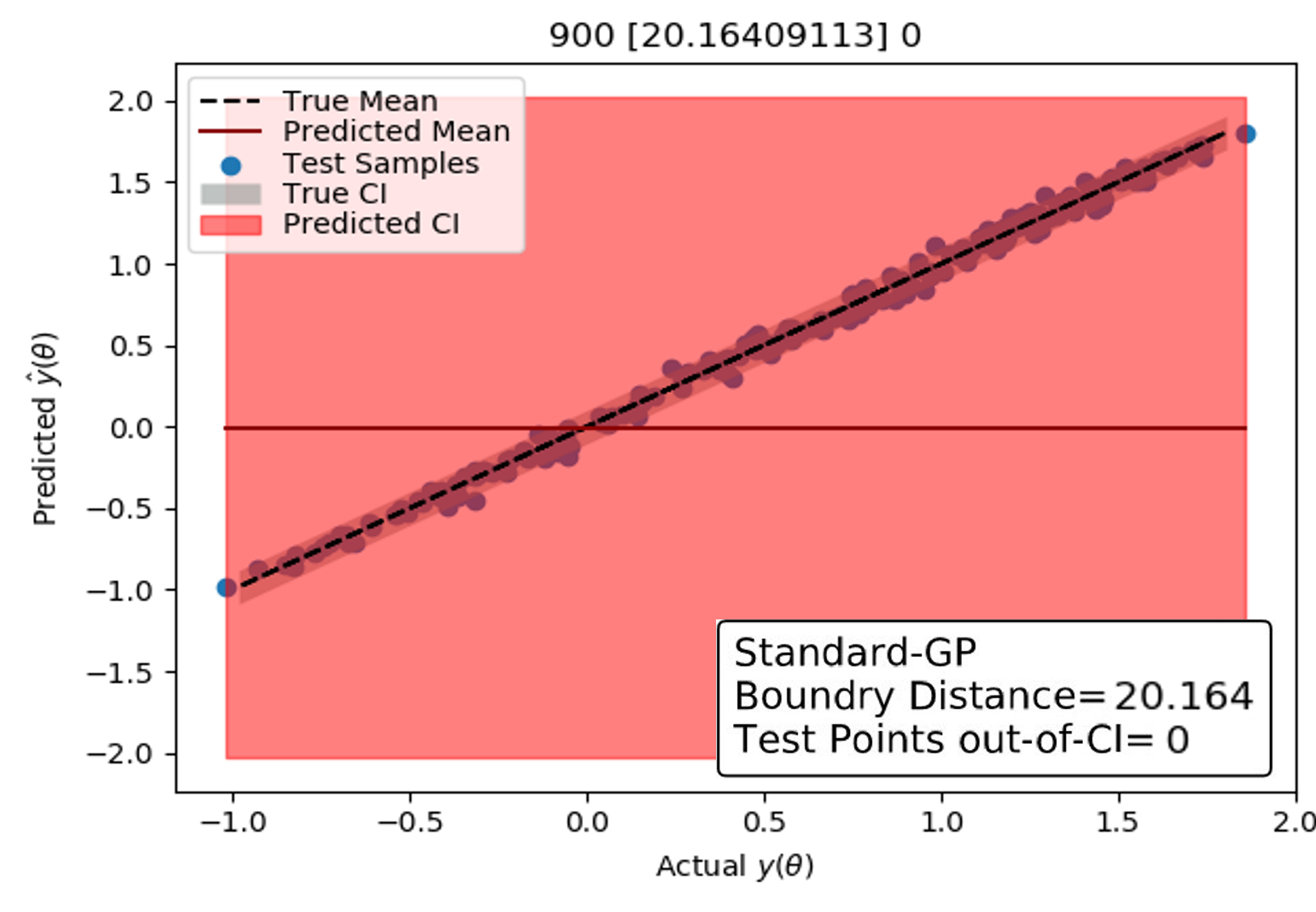} &
		\includegraphics[width=0.3\linewidth]{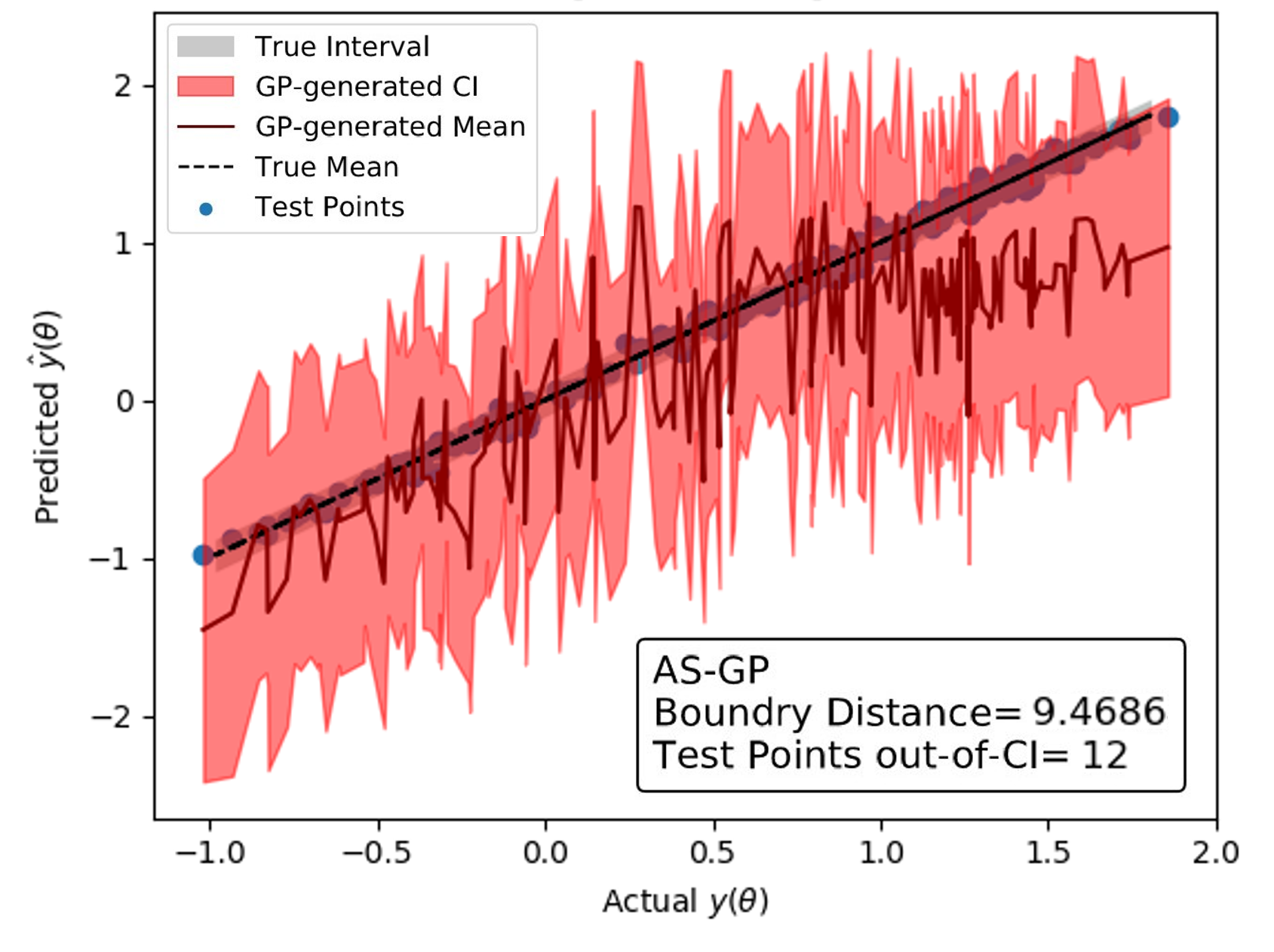} &
		\includegraphics[width=0.3\linewidth]{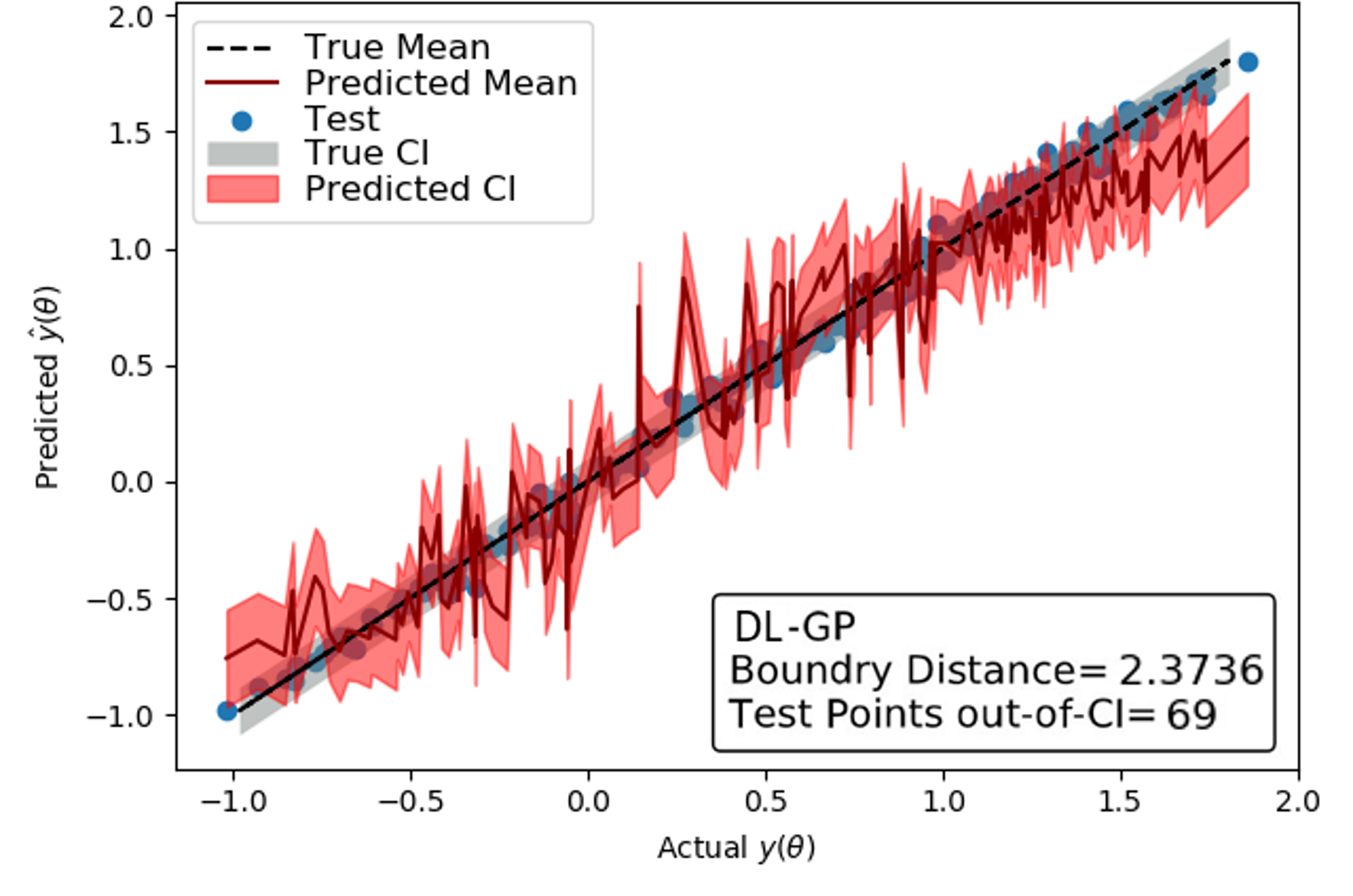} \\
		(a) $Original$ + $\mathcal{GP}_{0}$ & (b) $AS$ + $\mathcal{GP}_{0}$ & (c) $DL$ + $\mathcal{GP}_{0}$
	\end{tabular}
	\caption{Comparison of $n_{test}=200$ test predictions by a constructed GP based on each reduction method with the true values. Since the true function is stochastic, the true values are captured as follows: the black line indicates the correct position if the predicted value mean matches the true mean; the gray shaded region indicates the true function's $95\%$ Confidence Interval; and the blue points are the evaluations provided as test data. The red line represents the mean predicted value of the GP; the red shaded region indicates the GP's $95\%$ Confidence Interval.The GP is built based on only the training set $n_{train}=80$.}
	\label{fig:Ex2_c}
\end{figure}	

Again the difference between the reduction methods and the original algorithm are quite obvious, again highlighting the advantage of our projection-based approach. Additionally, the magnitude in difference in the uncertainty of the $AS$ + $\mathcal{GP}_0$ method, along with the repeated curvature seen in the $AS$ method's original prediction graph (Figure \ref{fig:Ex2_b}), demonstrate the importance for having a non-linear dimension reducer when encountering non-linear problems. 

\subsection{ABM Calibration}
We demonstrate our framework using the POLARIS ABM model for ground transportation. 
We demonstrate the methodology introduced in this paper by calibrating key parameters of the POLARIS ABM model of ground transportation in Bloomington, Illinois. All variables not being calibrated are assumed to be known. We generate an initial sample of size $N=16$, which is used to estimate the parameters of our dimensionality reduction maps, as well as the parameters of the mean function. The initial sample set was generated via Latin Hypercube. Then we apply BO with and without a dimensionality reduction pre-processing step. Specifically we evaluated the following dimensionality reduction configurations:

\begin{list}{$\bullet$}{}
	\item {\textbf{Original ($Original$)}}- no dimensionality reduction
	\item {\textbf{Active Subspace ($AS$)}} - the 2-dimensional subspace resulting from the Active Subspace dimensionality reduction pre-processing outlined in Appendix \ref{Ap:AS}
	\item {\textbf{Deep Learning Multi-layer Network ($DL$)}} - the 3-dimensional subspace resulting from the deep learning dimensionality reduction pre-processing.

\end{list}
Then we combine each of the dimensionality reduction configuration with one of the two GP profiles: a zero-mean GP $\mathcal{GP}_0$, and a deep learning multi-layered network-adjusted mean GP $\mathcal{GP}_{DL}$. Both instances use a Mat\'ern covariance function.

\begin{figure}
	\centering
	\begin{tabular}{cc}
		\includegraphics[width=0.3\linewidth]{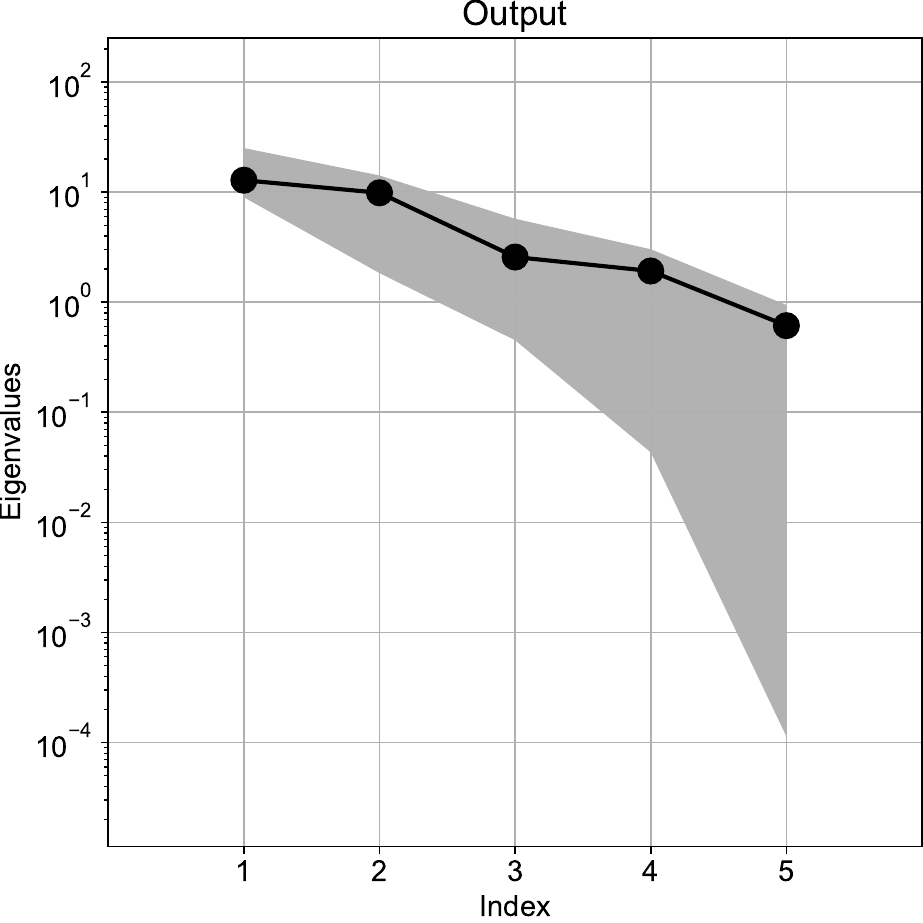} &
		\includegraphics[width=0.35\linewidth]{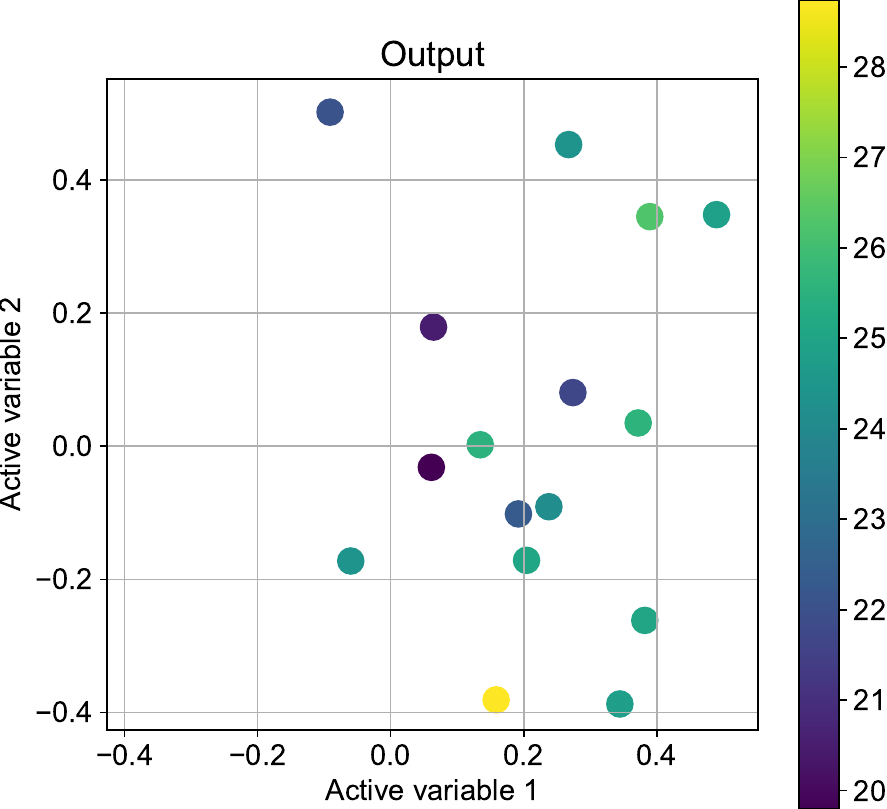} \\
		(a) Eigenvalues by Eigenvector & (b) $2$-D Subspace\\	
	\end{tabular}
	\caption{(a) Eigenvalues for each variable Eigenvector (black circles) and accompanying Bootstrap Interval (grey regions) found using $1000$ bootstrap replicates. (b) Training samples projected onto the $2$-D active subspace.}
	\label{fig:AS}
\end{figure}

The choice of latent space dimensionality for the $AS$ algorithm was determined from the elbow plot shown in Figure \ref{fig:AS}(a), which shows the eigenvalues and the 95\% confidence interval calculated from $1000$ bootstrap samples.

The gap between the second and third variables indicate the likely presence of a $2$-dimensional active subspace, which is supported by Figure \ref{fig:AS}(b). Notably, the bounds in the bootstrap intervals are quite large; however, given that the bootstrap intervals are most bias at low sample numbers, this outcome is not unexpected. 

A $3$-dimensional subspace was constructed for the $DL$ dimension reduction algorithm using the structure visualized in Figure \ref{fig:5_{DL}}.

\begin{figure}[H]\vspace*{4pt}
	\centering
	\includegraphics[width=.8\linewidth]{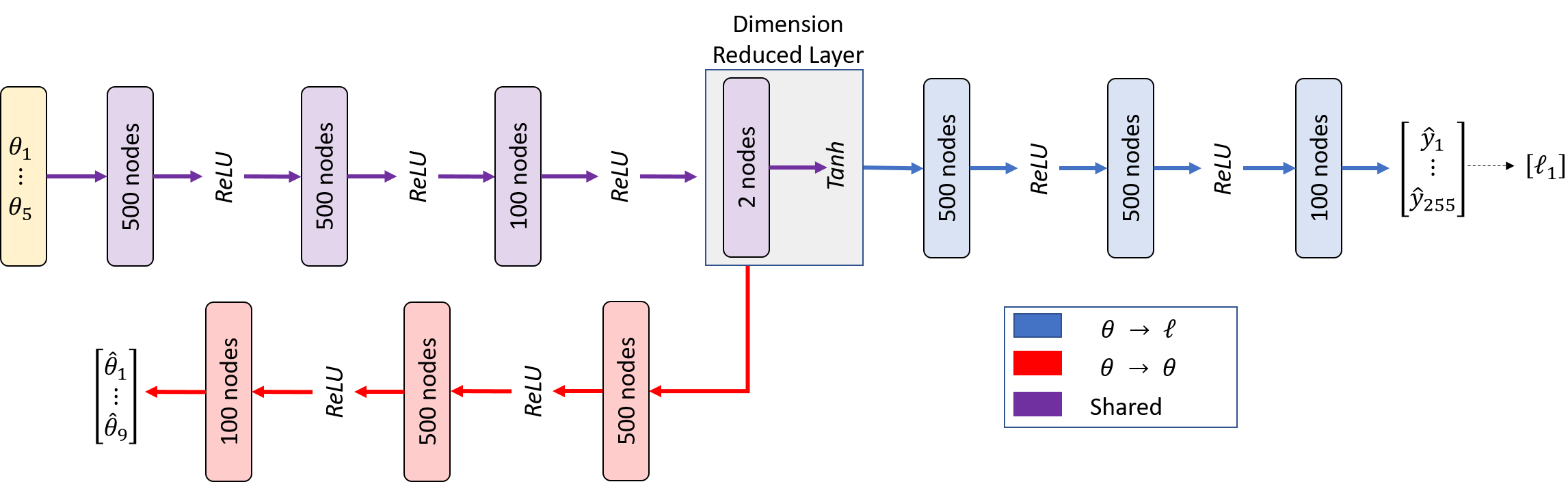}
	\caption{The implemented $DL$ structure for calibration. Rectangles represent the number of affine transformations applied to their layer inputs.}
	\label{fig:5_{DL}}
\end{figure}

\begin{figure}[H]\vspace*{4pt}
	\centering
	\includegraphics[width=.5\linewidth]{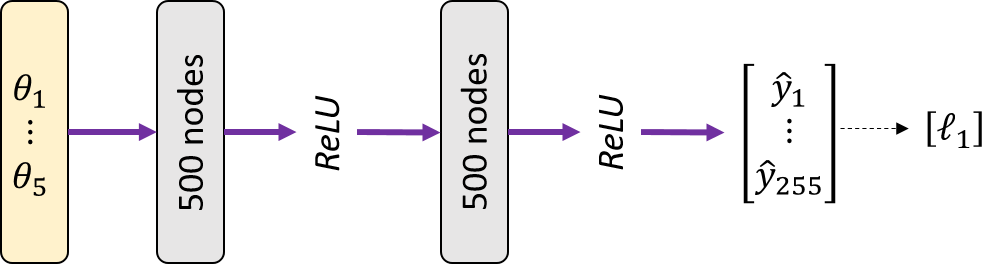}
	\caption{The implemented network structure of the adjusted mean function. The rectangles represent the affine transformations applied to their layer inputs.}
	\label{fig:5_MeanDL}
\end{figure}

A $ReLU$ function was applied universally with the exception of the reduced dimension layer, which utilized a $\tanh$ transformation to enforce a range boundary. The network was trained on $1000$ epochs with a training function $R(\theta)_{DL}$ using a penalty value of $P=200$ for decoding outside the original subspace bounds and a scalar $\lambda=0.005$ in order to encourage equal consideration between learning the decoding and encoding structures:
\begin{equation}
R_{DL}(\theta) = 0.005 \lVert \mathcal{L}-\hat{\mathcal{L}} \rVert_F + \lVert \theta-\hat{\theta} \rVert_F + 200\sum_{i=1}^{N}\left[\max[0,\hat{\theta_i}-\theta_{ui}]^2 + \max[0,\theta_{li}-\hat{\theta_i}]^2\right]
\end{equation}
Each subspace method, the same deep learner mean structure was used for the modified BO GP, depicted in \ref{fig:5_MeanDL}. 

A $ReLU$ function was applied universally and the network was trained on $800$ epochs with a learning rate of $0.01$; a simple MSE loss function was used along with an Expected Improvement (EI) utility criterion with batch-size of $2$. 

Now we apply our Bayesian Optimisation algorithm described in Section \ref{sec:BO} to the problem of calibrating the five parameters given in Table \ref{tab:params}. We compare the performance of BO without dimensionality reduction, and with zero mean ($Original$ + $\mathcal{GP}_0$), as well as with the non-linear mean function defined by a deep learning multi-layered network ($Original$ + $\mathcal{GP}_{DL}$). Further, we evaluate the effect of two dimensionality reduction techniques on the convergence of our algorithm. 

\begin{figure}[H]\vspace*{4pt}
	\centering
	\includegraphics[width=.7\linewidth]{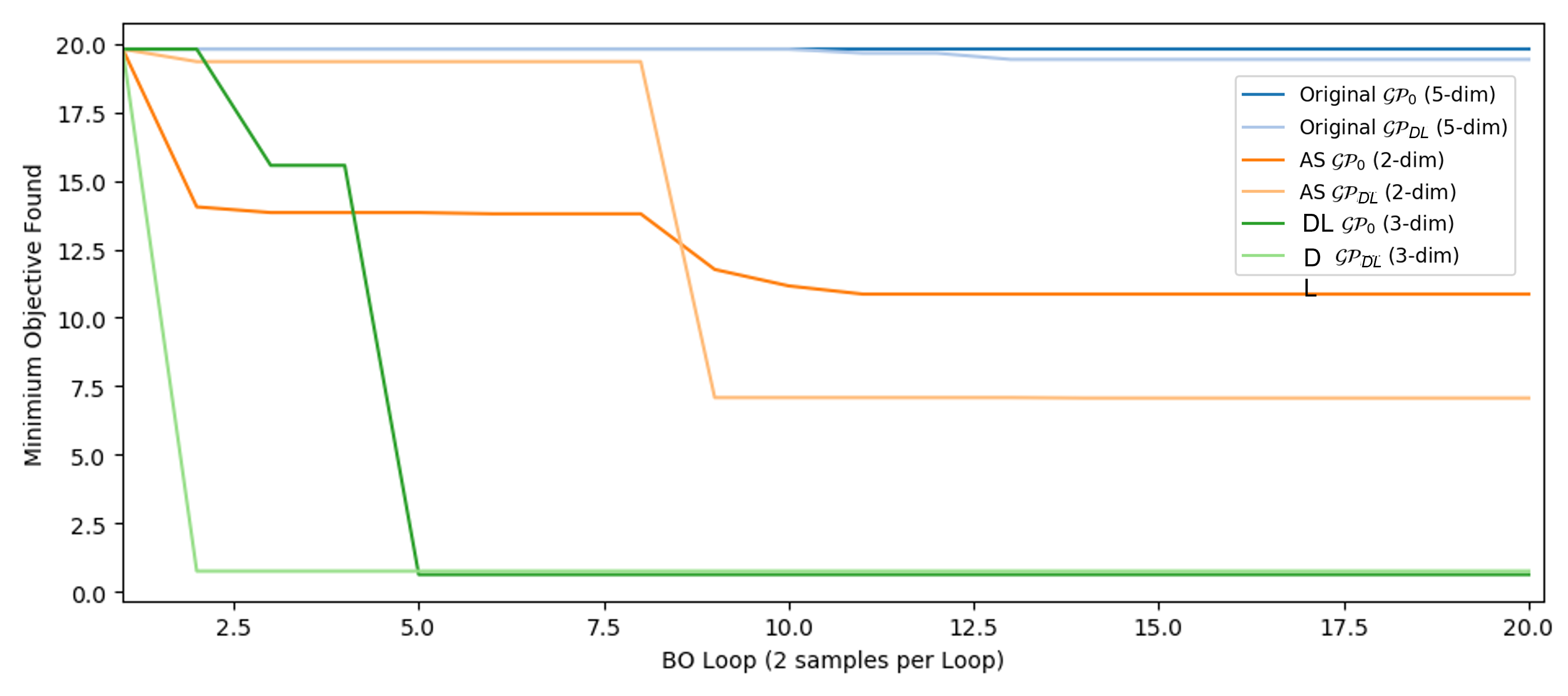} 
	\caption{Comparison of the minimum objective value found in $20$ BO iterations by each configuration during the calibration of $5$ dimensions.}
	\label{fig:5_results}
\end{figure}

Figure \ref{fig:5_results} shows the value of the objective function $\mathcal{L}$ for six different variations of our optimisation algorithm. 

Although vanilla BO is usually applicable to $5$-dimensional problems, the highly non-linear and heteroskedastic nature of our problem makes this BO approach impractical. The $Original$ + $\mathcal{GP}_0$ BO yields no improvements after $20$ iterations. When the non-zero DL mean was used $Original$ + $\mathcal{GP}_DL$, the  BO only manages an $2\%$ improvement after $12$ iterations. 

The $AS$ + $\mathcal{GP}_0$ configuration achieves a $30\%$ objective reduction within the first iterations; within $10$ iterations, it ultimately settles at an objective value of $10.86$ for a total improvement of $45\%$ and then stagnates. Adding a $\mathcal{GP}_{DL}$ mean improves on the $AS$ + $\mathcal{GP}_0$ result by $35\%$, leading to a final objective value of $7.07$ and a $65\%$ final improvement on the initial solution. However, it required 6 more iterations for $AS$ + $\mathcal{GP}_{DL}$ to achieve the improvement.

\begin{figure}[H]\vspace*{4pt}
	\centering
	\begin{tabular}{ccc}
		\includegraphics[width=0.3\linewidth]{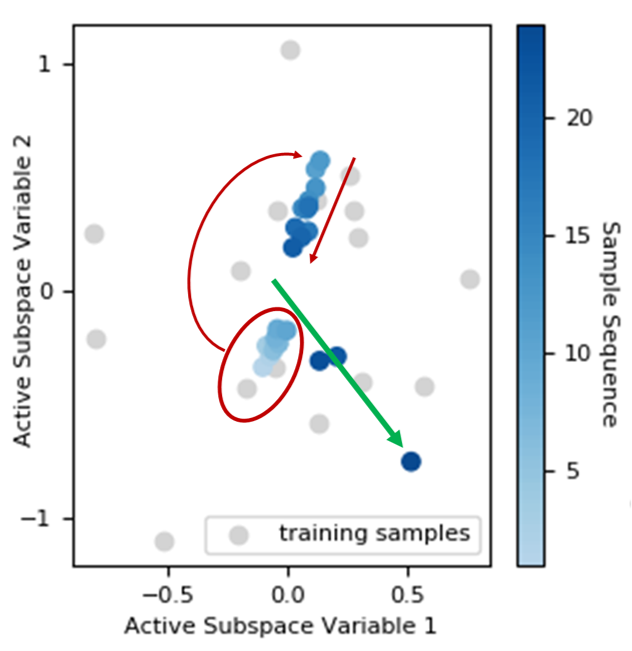} &
		\includegraphics[width=0.31\linewidth]{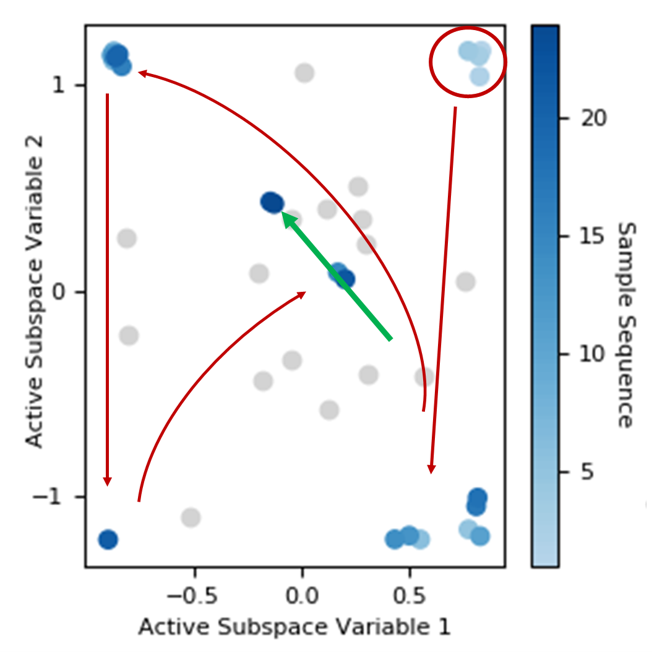} &
		\includegraphics[width=0.3\linewidth]{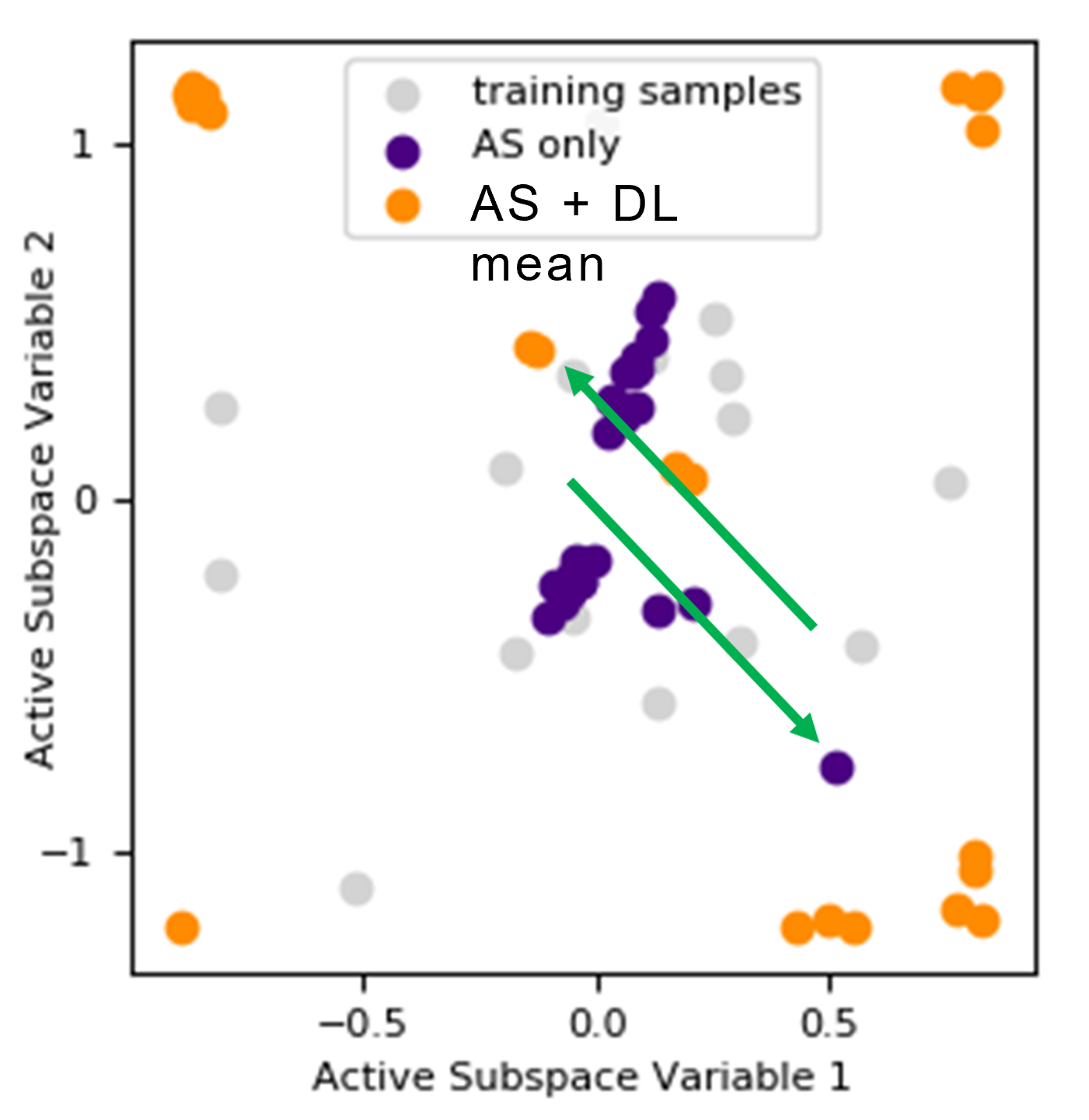}\\
		(a) AS + $\mathcal{GP}_0$ & (b) AS + $\mathcal{GP}_{DL}$ & (c) AS + $\mathcal{GP}_0$ vs AS + $\mathcal{GP}_{DL}$ \\	
	\end{tabular}
	\caption{Samples in latent subspace that were used by the search algorithm. The grey dots correspond to the initial $16$ samples and the $40$ colored dots were sequentially recommended in over the $20$ BO iterations. The intensity of the blue color corresponds to the order in which the samples were recommended. The red circles are initial recommendations, the red arrows show the sampling directions, and the green arrow highlights the final trajectory.
	(a) $AS$ + $\mathcal{GP}_0$  (b) $AS$ + $\mathcal{GP}_{DL}$ (c) overlays (a) in purple and (b) in orange to highlight that the final trajectories are parallel but offset.}
	\label{fig:as_vs_mean}
\end{figure}

Figure \ref{fig:as_vs_mean} shows that the use of the $\mathcal{GP}_{DL}$ leads to more exploration by the search algorithm. While, $AS$ + $\mathcal{GP}_0$ primarily employed exploitation of the initial training set, recommending samplings within a tight cluster about the majority of the training points in the center of the subspace, the $AS$ + $\mathcal{GP}_{DL}$ explored the unknown portions of the search space. 

\begin{figure}[H]\vspace*{4pt}
	\centering
	\begin{tabular}{ccc}
		\includegraphics[width=0.31\linewidth]{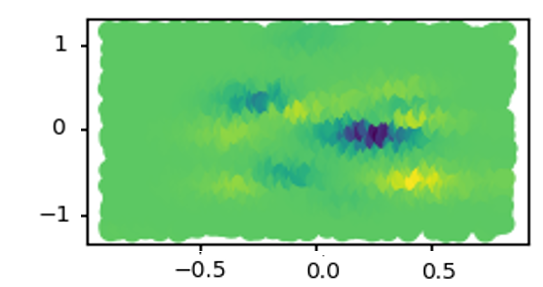} &
		\includegraphics[width=0.31\linewidth]{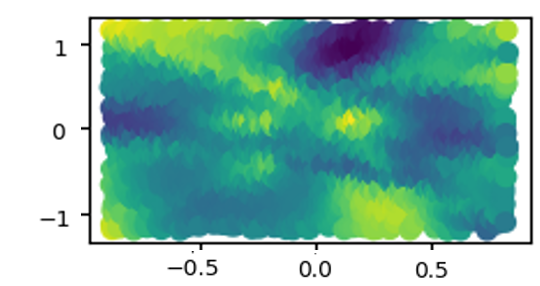} &
		\multirow{5}{*}[0.7in]{
		\includegraphics[width=.1\textwidth]{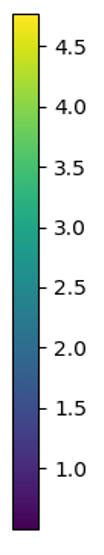}} \\
		(a) $AS$ + $\mathcal{GP}_0$ $\mu_\mathrm{post}$ & (b) $AS$ + $\mathcal{GP}_{DL}$ $\mu_\mathrm{post}$ & \\
		\includegraphics[width=0.31\linewidth]{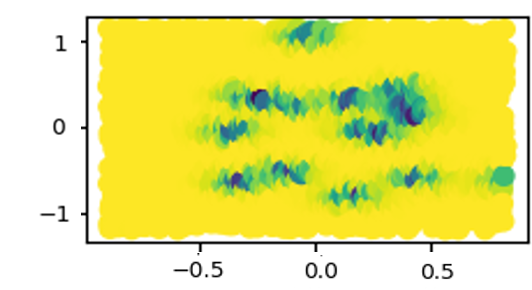} &
		\includegraphics[width=0.31\linewidth]{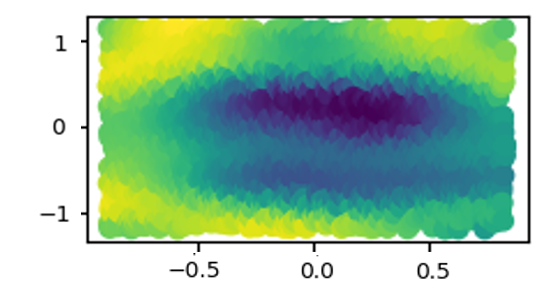} & \\
		(c) $AS$ + $\mathcal{GP}_0$ $K_\mathrm{post}$& (d) AS + $\mathcal{GP}_{DL}$ $K_\mathrm{post}$ & \\
		\includegraphics[width=0.31\linewidth]{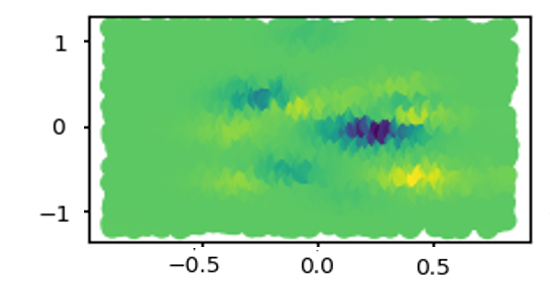} &
		\includegraphics[width=0.3\linewidth]{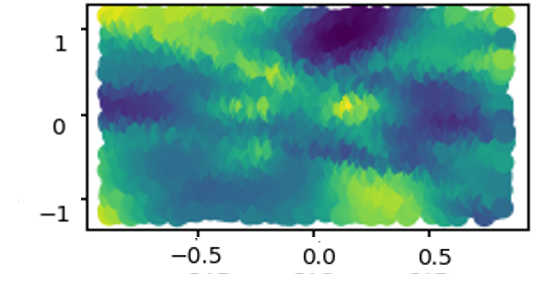} & \\
		(e) $AS$ + $\mathcal{GP}_0$ EI Value & (f) $AS$ + $\mathcal{GP}_{DL}$ EI Value & \\
	\end{tabular}
	\caption{Expected Improvement components and acquisition values for candidate points in the $AS$ statespace during an interval of the BO calibration for $AS$ + $\mathcal{GP}_0$ (a)(c)(e) and $AS$ + $\mathcal{GP}_{DL}$ (b)(d)(f). (a) and (b) show the predicted mean reduction component, (c) and (d) show the predicted standard deviation component, and (e) and (f) show the final acquisition calculation result. The acquisition function in both settings rely heavily on the mean reduction component despite their diverging recommendation patterns.}
	\label{fig:EI_diff}
\end{figure}

Typically, the change in behavior would imply that the Expected Improvement (EI) acquisition function faced higher variance due to the DL mean as, broadly, EI encourages exploitation when a strong reduction in the mean function is predicted and exploration when the variance is high. However, as shown in Figure \ref{fig:EI_diff}, this is not the case. 

While both $\mathcal{GP}_0$ and $\mathcal{GP}_{DL}$ acquisition values clearly rely heavily on the potential in mean reduction, it is the landscape of the predicted mean and uncertainty which shifts significantly. This outcome is likely due to the newly reflected magnitude of non-linear behaviors previously unaddressed by the $AS$ + $\mathcal{GP}_0$ configuration. Figure \ref{fig:as_vs_mean}(c) shows this to indeed be the case, where the $AS$ + $\mathcal{GP}_{DL}$ required an offset to locate improved recommendations.


The largest advancement was produced by the $DL$ + $\mathcal{GP}_{0}$ configuration at $96.85\%$ with a final objective value of $0.625$; furthermore, this achievement was located within just $3$ iterations. The addition of the $\mathcal{GP}_{DL}$ netted a similar reduction but accelerated the discovery to within the first iteration. Table \ref{tbl:5_summary} provides summary findings.

\begin{table}[H]\vspace*{4pt}
	\centering
	\includegraphics[width=.65\linewidth]{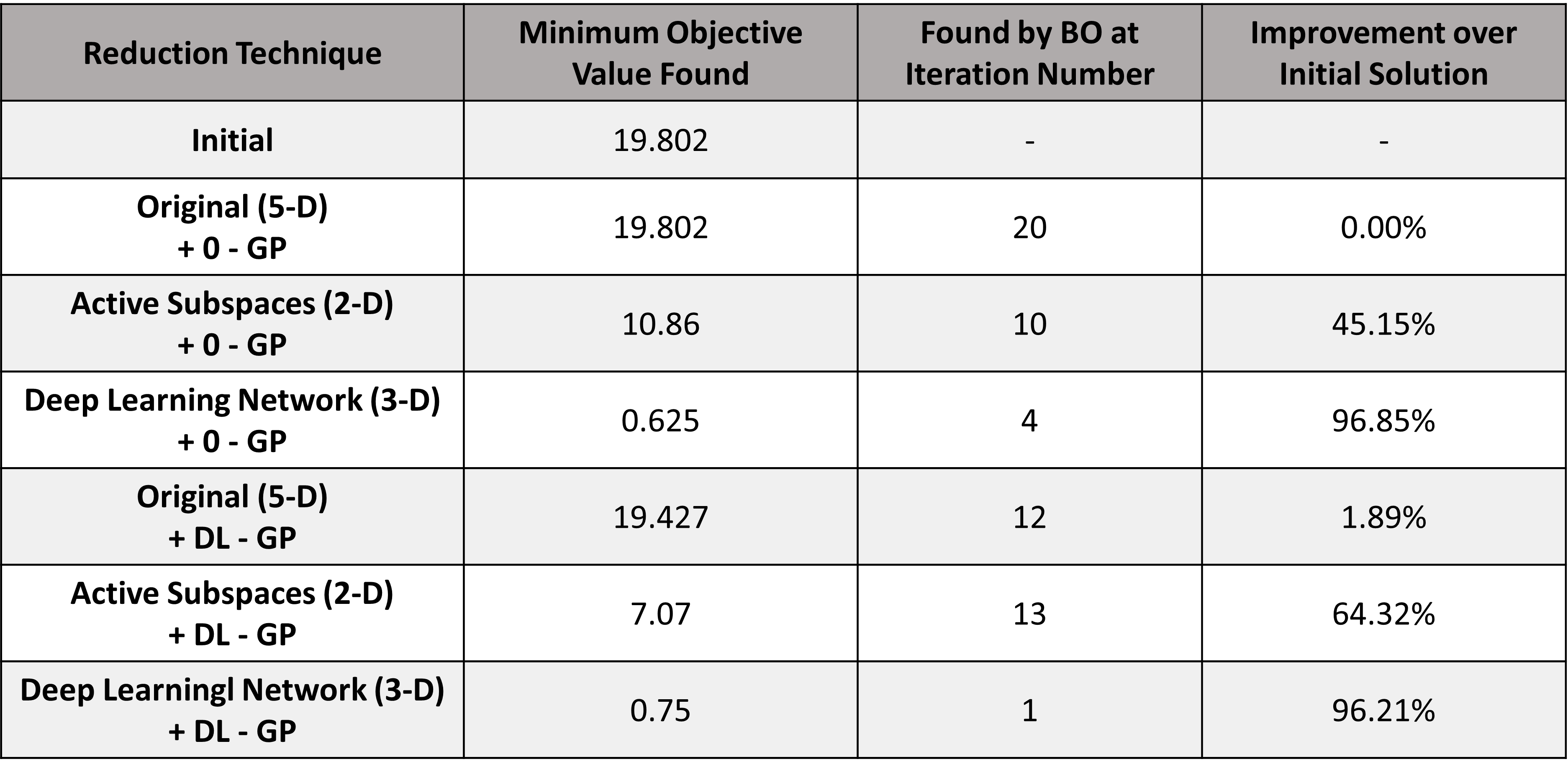}
	\caption{Comparison by configuration of their best solutions found during a maximum of $20$ BO iterations. Each iteration produced $2$ recommendations for evaluation.}
	\label{tbl:5_summary}
\end{table}

\subsection{Predictions}

In this section, we review the improvements made to the simulation outputs as a result of the calibration process. We treat the best of the initial $16$ LHS samples as a baseline to compare with the $DL$ + $\mathcal{GP}_0$ solution. 

\begin{figure}[H]
	\centering
	\begin{tabular}{ccc}
		\includegraphics[width=0.3\linewidth]{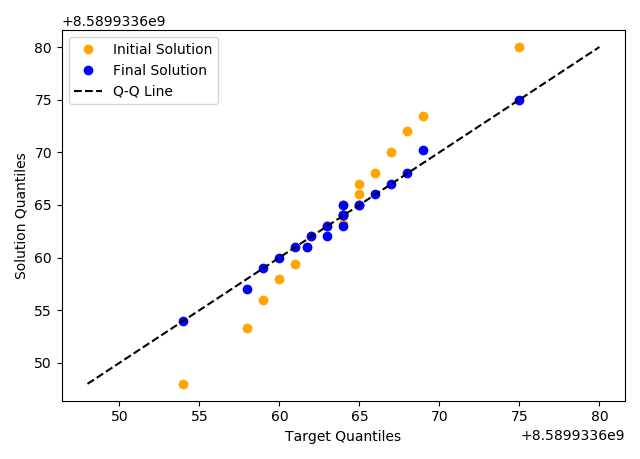} &
		\includegraphics[width=0.3\linewidth]{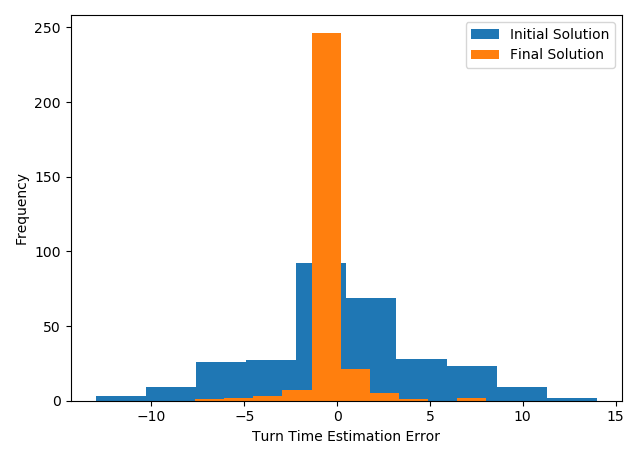} &
		\includegraphics[width=0.3\linewidth]{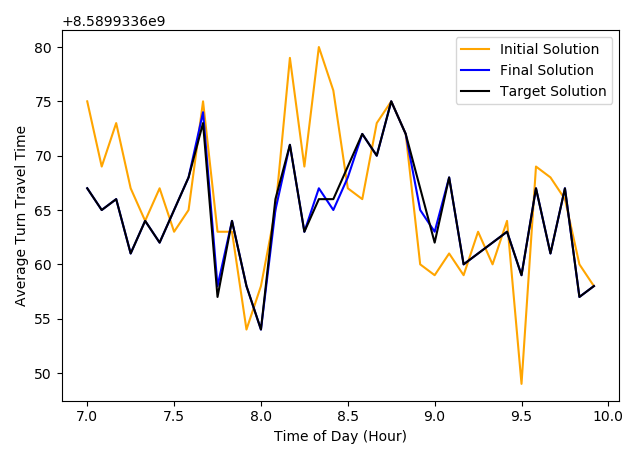}\\
		(a) QQ-Plot & (b) Error Histogram & (c) Comparison 7am - 10am \\	
	\end{tabular}
	\caption{Comparison of the targeted outputs with the initial solution found during the $16$ initial sample set and the final solution found by $DL$ + $\mathcal{GP}_0$ during the $20$ iterations of BO. (a) compares each solution's Quantiles to the targeted solution's Quantiles. The black, dashed line indicates where the ideal solution should lie. (b) captures the distribution of output errors between the solutions and the targeted solution. (c) shows the average turn times around and at rush hour (7 am - 10 pm)}
	\label{fig:improvements}
\end{figure}

Figure \ref{fig:improvements}(a) takes a non-parametric approach to compare the solutions in respect to the targeted outputs via a Quantile-Quantile (Q-Q) plot. The plot shows the initial solution aligning in the central quantiles but curving off on either side while the calibrated solution remains in consistent alignment. This indicates that the initial solution was heavy-tailed, or possessed more extreme values than the true solution, and that the calibration process was able to correct the behavior. Figure \ref{fig:improvements}(b) further highlights this accomplishment. The calibrated solution is shown to not only improve the frequency of correct outputs but tighten the initial solution's extreme deviations.


\section{Discussion}\label{sec:discussion}
We provided a Bayesian optimization framework for complex transportation simulators. We have shown that Bayesian optimization algorithms, when combined with dimensionality reduction techniques, provide a good default option for the calibration of transportation ABMs. The Bayesian approach provides several advantages over other black-box optimization techniques: First, the conditional distribution over the output, provides an efficient approach to maximizing a limited sampling budget (sequential design of computational experiment). Second, the posterior distribution allows for sensitivity analysis and uncertainty quantification to be performed. Although Bayesian optimization has been proposed before for simulation-based problems, our approach, which combines it with dimensionality reduction techniques, allows for solving real-life high-dimensional problems. Our parallel batching strategy and high performance computing framework allow for these approaches to be used by both researchers and transportation practitioners. Future research should consider 'opening the box' and shifting the algorithm from recommending future samples based on an aggregated loss function, which averages out the variance across time, to the potential improvements by individual time periods. The model could potentially identify the most important contributors to the acquisition function in this manner and encourage the algorithm to direct its attention to those areas for concentrated improvement.


\appendix
\section{Active Subspaces}\label{Ap:AS}

Active Subspaces (AS) identifies a linear transformation of the subspace by constructing a set of uncorrelated eigenvectors, known as  Principal Components (PCs), defining the "active" subspace. These orthogonal directions, denoted as $W$, capture the average variability of the $\theta$-to-$\mathcal{L}$ relationship via a differential function $\mathcal{L} \approx f(\theta): \mathbb{R}^d \to \mathbb{R}$:

\begin{align}
\max_{\lVert W \rVert = 1} & \mathbb{E}\left[\lVert \mathcal{L}(\theta W) \rVert_0\right]\\
& \mathbb{E}\left[W^T\theta^T\mathcal{L}^T\mathcal{L}\theta W\right] \\
& W^T\mathbb{E}\left[cov^2(\theta V,mathcal{L})\right]W\\
& W^TCW
\end{align}

Given a bounded probability density function $\rho(\theta) \in \mathbb{R}^d$ on $f(\theta)$, AS first decomposes the eigenvalue-eigenvector representation on the gradient $\nabla_\theta$:

\begin{equation}
C = \int (\nabla_{\theta}f)(\nabla_{\theta}f)^T\rho(\theta)d\theta = \mathcal{W}\Lambda \mathcal{W}^T
\end{equation}

where $C$ is a sum of semi-positive definite rank-one matrices, $\mathcal{W}$ is the matrix of eigenvectors, and $\Lambda$ is the diagonal matrix of eigenvalues in decreasing order. 

In cases where the gradient is unknown, as with most simulations, $C$ can instead be approximated using an observed set of $n$ inputs sampled from $\rho(\theta)$ via Monte Carlo:

\begin{equation}
\hat{C}=\frac{1}{n}\sum_{i=1}^{n} (\hat{\nabla}_\theta f_i)(\hat{\nabla}_\theta f_i)^{T} = \hat{\mathcal{W}} \hat{\Lambda} \hat{\mathcal{W}}^{T}
\end{equation}

where $\hat{\nabla}_{\theta}$ is the observed gradient or an estimated gradient, such as a local or global linear regression models: 
\begin{center}
	\begin{equation}
	\begin{array}{c}
	{f}_i\approx {\hat{\beta}}_{0}+\hat{\beta}^{T}\theta_i\\
	\nabla_\theta f_i = \hat{\beta} 
	\end{array}
	\end{equation}
\end{center} 

Once determined, the PCs are plotted on a log-scale and a dramatic drop in the eigenvalue space, documented as a \textit{gap}, is sought~\footnote{if no gap can be found, compiling larger sets of eigenvalues or sampling more within the current eigenvalue framework to increase the eigenvalue accuracy is suggested}:

\begin{figure}[H]
	\centering
	\includegraphics[width=0.3\linewidth]{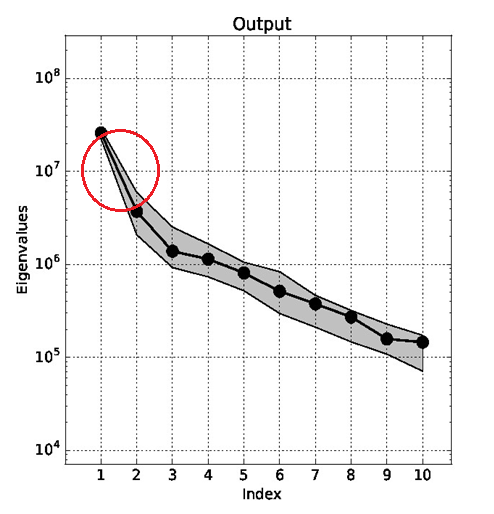}	
	\caption{The largest eigenvalue gap is designated by a red circle}
	\label{fig:activesubspaces}
\end{figure}

The dimensions $q<d$ to the left of the largest eigenvalue gap is designated the "active" subspace $v=\hat{\mathcal{W}}_1^T\theta$; the remaining $d-n$ directions define the "inactive" subspace $z=\hat{\mathcal{W}}_2^T\theta$.
\begin{equation}
\theta = \hat{\mathcal{W}}\hat{\mathcal{W}}^T\theta = \hat{\mathcal{W}}_1 v + \hat{\mathcal{W}}_2 z
\end{equation}

The variables of $z$ are then fixed at nominal values and the active set $\hat{\mathcal{W}}_1$ subsequently becomes the approximate representation of $\Theta$:

\begin{equation}
\begin{split}
f(\theta) &\approx \int f\left(\hat{\mathcal{W}}_1v + \hat{\mathcal{W}}_2 z\right)p\left(v \mid z\right) dv \\
&\approx g\left(v\right) = g\left(\hat{\mathcal{W}}_1^T\Theta \right)
\end{split}
\end{equation}

For more information behind the derivation of this procedure and its variations, see \cite{constantine_active_2015}.

Like PCA, AS will require a subsequent, separate regression from the subspace to the output variables $v \to f(\theta)$:
\begin{align}
f(\theta) &=\beta \theta +\epsilon\\
&\approx \beta v +\epsilon \equiv \beta \mathcal{W}_1^T\theta +\epsilon
\end{align}

Active Subspaces will provide results closer in line with Partial Least Squares (PLS). This is because, through leveraging the derivatives, the method actively prioritizes those directions which are also relevant to the input-output relationship ultimately being sought. However, the inversion of the lower-dimensional subspace to recover the original variables is complicated by the inactive subspace. For situations in which the inactive eigenvalues are exactly zero, $f(\theta)$ can be reconstructed without regard to the inactive subspace by setting $z=0$:

\begin{equation}
f\left(\theta\right)\approx f(\hat{\mathcal{W}}_{1}v) = f (\hat{\mathcal{W}}_{1}\hat{\mathcal{W}}_{1}^{T}\theta)
\end{equation}

When the inactive eigenvalues are small but not zero, decoding should be constructed by optimizing over the inactive subspace $v$ such that:

\begin{equation}
\underset{v \in V}{\text{minimize}} \left[\begin{array}{cc}
\underset{z \in Z}{\text{minimum}}  & f\left(\hat{\mathcal{W}}_{1}v+\hat{\mathcal{W}}_{2}z\right) \\ \text{s.t.} &  \Theta_{lb}-\hat{\mathcal{W}}_{1}v \le \hat{\mathcal{W}}_{2}z \le \Theta_{ub} - \hat{\mathcal{W}}_{1}v
\end{array}\right]
\end{equation}

where $\Theta_{lb}$ and $\Theta_{ub}$ are the lower and upper bounds of the original $\Theta$ state-space, respectively.

\bibliography{IEEEfull_Bibtex}



\end{document}